\documentclass[%
reprint, 
superscriptaddress,
nofootinbib,
 amsmath,amssymb,
 aps,
showkeys
]{revtex4-2}

\usepackage{graphicx}% Include figure files
\graphicspath{{main_figures/}}
\usepackage{dcolumn}% Align table columns on decimal point
\usepackage{bm}% bold math
\usepackage{hyperref}% add hypertext capabilities
\usepackage{siunitx}

\begin{document}

\preprint{}

%\title{Large area multiwavelength and polarization-insensitive metalenses via inverse designed multilayer Huygens’ metasurfaces}% Force line breaks with \\

\title{Large-area multiwavelength and polarization-insensitive metalenses via multilayer Huygens’ metasurfaces}% Force line breaks with \\

\author{Joshua Jordaan}
    \affiliation{Institute of Solid State Physics, Friedrich Schiller University Jena, 07743 Jena, Germany}%Lines break automatically or can be forced with \\
    \affiliation{Institute of Applied Physics, Abbe Center of Photonics, Friedrich Schiller University Jena, 07745 Jena, Germany}
    \affiliation{ARC Centre of Excellence for Transformative Meta-Optical Systems (TMOS), Research School of Physics, Australian National University, Canberra ACT 2601, Australia}
 
\author{Alexander Minovich}%
    \affiliation{Institute of Solid State Physics, Friedrich Schiller University Jena, 07743 Jena, Germany}%Lines break automatically or can be forced with \\
    \affiliation{Institute of Applied Physics, Abbe Center of Photonics, Friedrich Schiller University Jena, 07745 Jena, Germany}

\author{Dragomir Neshev}
    \affiliation{ARC Centre of Excellence for Transformative Meta-Optical Systems (TMOS), Research School of Physics, Australian National University, Canberra ACT 2601, Australia}

% \author{Thomas Pertsch}
%     % \affiliation{Institute of Applied Physics, Abbe Center of Photonics, Friedrich Schiller University Jena, Albert-Einstein-Straße 15, Jena, 07745, Germany}%Lines break automatically or can be forced with \\
%     \affiliation{Institute of Applied Physics, Abbe Center of Photonics, Friedrich Schiller University Jena, 07745 Jena, Germany}
%     % \affiliation{Max Planck School of Photonics, Hans-Knöll-Straße 1, Jena, 07745, Germany}
%     \affiliation{Max Planck School of Photonics, 07745 Jena, Germany}
%     % \affiliation{Fraunhofer Institute for Applied Optics and Precision Engineering IOF, Albert-Einstein-Straße 7, Jena, 07745, Germany}
%     \affiliation{Fraunhofer Institute for Applied Optics and Precision Engineering IOF, 07745 Jena, Germany}

\author{Isabelle Staude}
    \email{isabelle.staude@uni-jena.de}
    \affiliation{Institute of Solid State Physics, Friedrich Schiller University Jena, 07743 Jena, Germany}%Lines break automatically or can be forced with \\
    \affiliation{Institute of Applied Physics, Abbe Center of Photonics, Friedrich Schiller University Jena, 07745 Jena, Germany}
    
\date{\today}

\begin{abstract}
\begin{center}
    \centering
    \includegraphics[width=0.4\textwidth]{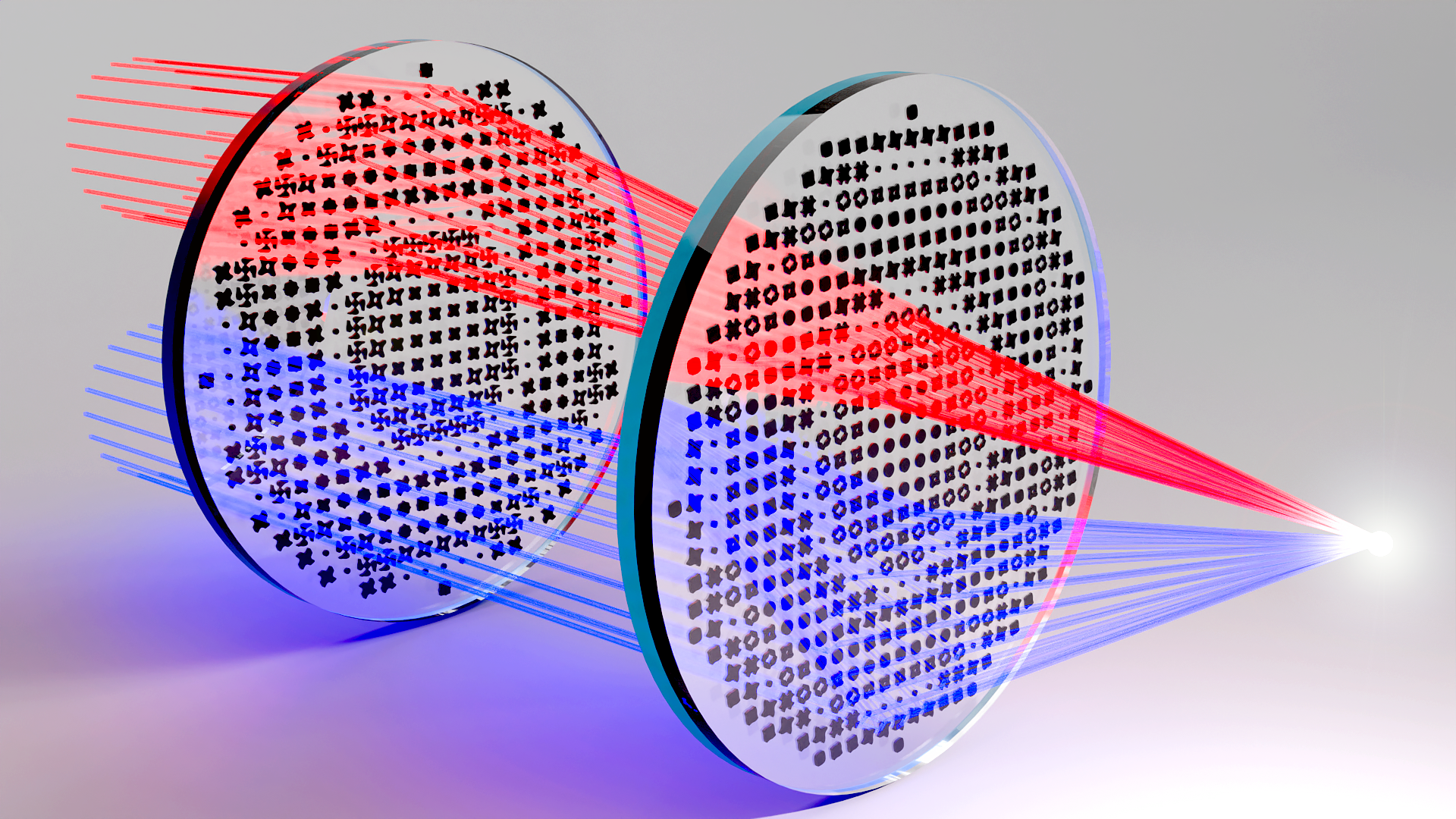}
\end{center}

Metalenses, advanced nanostructured alternatives to conventional lenses, significantly enhance the compactness and functionality of optical systems. Despite progress in monochromatic applications, scaling metalenses to centimeter-sized apertures for broadband or multiwavelength use remains challenging due to limitations in achieving the necessary group delay (GD) with current materials. In this study, we introduce a novel multiwavelength, polarization-insensitive metalens design operating in the near-infrared (NIR). Our approach employs multiple Huygens' metasurface layers, each optimized to modulate a specific wavelength while maintaining high transmittance and minimal phase disturbance at other wavelengths. We demonstrate a metalens operating at 2000 and 2340 nm with a numerical aperture (NA) of 0.11. In simulation, the metalens achieves a normalized modulation transfer function (MTF) that is nearly exactly diffraction-limited. The absolute focusing efficiencies are 65\% and 56\%, corresponding to relative efficiencies of 76\% and 65\% compared to an ideal lens of the same dimensions. The meta-atoms are designed using an inverse shape-optimization method that ensures polarization insensitivity and high tolerance to layer misalignment. The proposed design is compatible with large-area nanofabrication techniques, allowing metasurface layers to be fabricated individually and assembled into the final device. This innovative approach is also generalizable to any arbitrary multiwavelength phase profile beyond that of simple lensing.
%\begin{description}
%\item[keywords]
%\end{description}
\end{abstract}

%Use showkeys class option if keyword
                              %display desired

\maketitle

%\tableofcontents

\section{Introduction}\label{sec:intro}

\begin{figure*}[]
\includegraphics{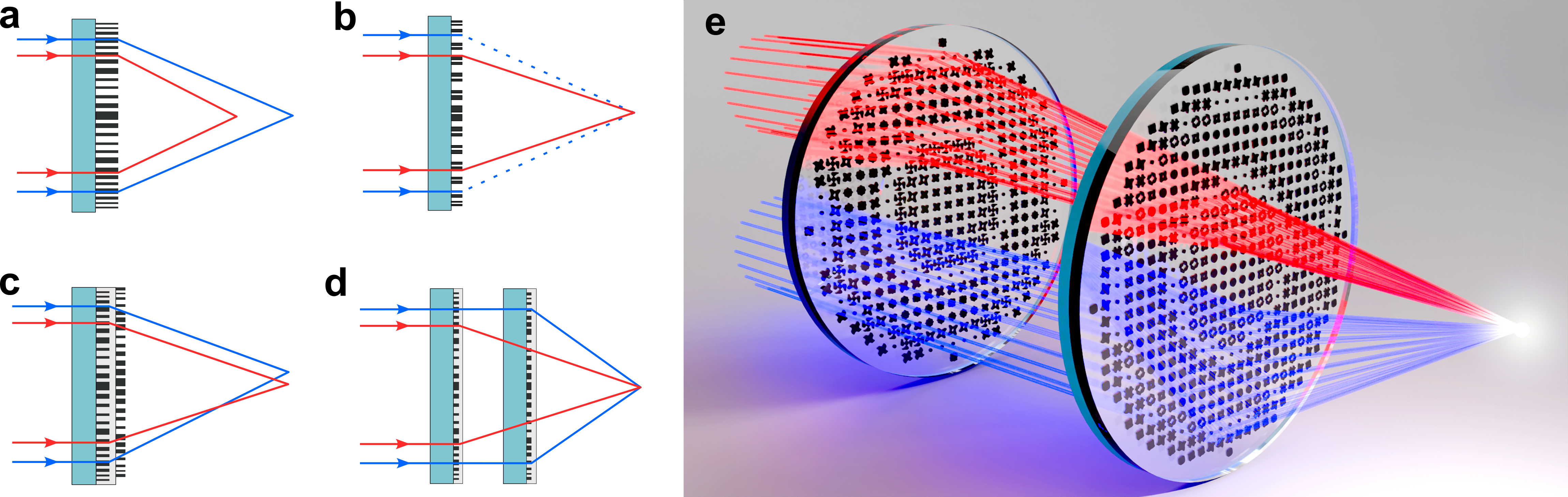}
\caption{\label{fig:schem and render} Comparison of different polarization-insensitive multiwavelength design schemes to create large-area metalenses. (\textbf{a-c}) Schematics of the main classes of current approaches illustrating their drawbacks: (\textbf{a}) dispersion engineering , (\textbf{b}) interleaving and (\textbf{c}) multilayer unitcells. (\textbf{d,e}) Schematic (\textbf{d}) and artistic render (\textbf{e}) showing a multilayer Huygens' metalens design. }
\end{figure*}
Metalenses, composed of flat arrays of subwavelength-sized scatterers known as meta-atoms, have rapidly emerged as a cutting-edge alternative to traditional diffractive and refractive lenses. By intelligently designing these scattering elements, metalenses offer unparalleled subwavelength control over the polarization, amplitude, phase, and frequency of incident light \cite{neshevEnablingSmartVision2023}. This advanced control can drastically enhance the efficiency and compactness of next-generation optical systems \cite{kuznetsovRoadmapOpticalMetasurfaces2024}. In the case of monochromatic operation, enormous progress has been made with high performing devices already demonstrated for many different wavelengths and up to tens of centimeters in size \cite{kimScalableManufacturingHighindex2023,sheLargeAreaMetalenses2018,parkAllGlassMassProducibleLargeDiameter2022}. However, a major limitation in achieving these aims for broadband or multiwavelength metalenses is the difficulty in scaling to the centimeter sized apertures which are critical for practical devices. This difficulty arises because the maximum required linear phase-dispersion, or group delay (GD) that needs to be applied to an incident field for ideal achromatic focusing, is proportional to the lens diameter \cite{presuttiFocusingBandwidthAchromatic2020, engelbergAchromaticFlatLens2021}. The restricted GD values that are in practice achievable in single layer nanostructured metalenses with current dielectrics places an upper bound on the diameter, which is well below that of the centimeter sizes needed in many real-world applications (see discussion in Supplementary S-I). This leads to significant chromatic aberration as the diameter is increased (Figure~\ref{fig:schem and render}a). To overcome this limitation, a recent study introduced the concept of a multizone dispersion-engineered metalens in which the applied GD is wrapped into discrete zones bounded by the attainable maximum value of the meta-atoms \cite{liMetaopticsAchievesRGBachromatic2021}. Each zone then achieves achromatic focusing independently, and phase offsets between zones are optimized such that they interfere constructively at the focal plane over the range of design wavelengths. A key limitation of this approach is that it necessitated the use of the geometric phase to independently control the phase and GD over the surface, such that the focusing response is polarization-sensitive. For systems operating with unpolarized sources, a polarization-sensitive response effectively halves the focusing efficiency and introduces unfocussed stray light, which detracts from system performance. Although a truly large-area, broadband and polarization insensitive metalens platform has yet to be demonstrated, progress has been made in the case of multiwavelength devices. Here, the GD constraint is relaxed and focusing scalable to large-areas can be achieved simply by applying a lens phase profile at each wavelength of interest utilizing an arbitrary phase-dispersion response. As most artificial sources operate at discrete wavelengths, multiwavelength rather than achromatic operation also does not significantly detract from many potential applications. Researchers have proposed several multiwavelength polarization-insensitive designs that are applicable to large-areas. These include annular interference \cite{caiUltrathinTransmissiveMetasurfaces2019,wangAchromaticFocusingEffect2021,avayuCompositeFunctionalMetasurfaces2017}, multilayer meta-atoms \cite{fengRGBAchromaticMetalens2022,zhouMultilayerNoninteractingDielectric2018}, spatial interleaving \cite{baekHighNumericalAperture2022,baekMultiwavelengthMetalensSpatial2021,arbabiMultiwavelengthMetasurfacesSpatial2016} and multiresonant \cite{arbabiMultiwavelengthPolarizationinsensitiveLenses2016,shiSingleLayerMetasurfaceControllable2018} approaches. For the annular interference based designs, the maximum expected efficiency is low and bounded by that of an ideal binary Fresnel zone plate at 10\% \cite{caiUltrathinTransmissiveMetasurfaces2019}. The multilayer meta-atom designs (Figure~\ref{fig:schem and render}c) in simulation are able to reach above 60\% focusing efficiencies for a numerical aperture (NA) of 0.2 but are heavily impacted by fabrication imperfections that vary the interlayer spacing and alignment \cite{zhouMultilayerNoninteractingDielectric2018}. The spatially interleaved and multiresonant designs (Figure~\ref{fig:schem and render}b) tend to suffer from a large variation in focusing performance over the design wavelengths, with generally below 40\% minimum focusing efficiency even in simulation. This effect can be attributed to a combination of unaccounted for nearfield interactions as a consequence of breaking the validity of the locally periodic approximation (LPA) and sparser spatial phase sampling at the shortest wavelengths \cite{baekHighNumericalAperture2022,arbabiMultiwavelengthPolarizationinsensitiveLenses2016}. To overcome these limitations, a recent study in Ref. \cite{liInverseDesignEnables2022} used an advanced large-area LPA based inverse design scheme to generate red, green and blue (RGB) polarization insensitive centimeter scale visible light metalenses with almost identical focusing efficiencies over the operating wavelengths. However, the maximum achieved focusing efficiencies in simulation were still low at $\sim$24\% due to the use of cross-polarization converting meta-atoms, which have characteristically lower efficiencies than isotropic meta-atom platforms.

In this work, we present a novel polarization-insensitive multiwavelength metalens design operating within the near-infrared (NIR). The design consists of multiple Huygens' metasurface layers, where each layer modulates only a specific wavelength, while achieving high transmittance and low phase disturbance at alternative wavelengths. This approach effectively circumvents the sparse phase sampling of spatial interleaving designs and minimizes the difficulty in maintaining the validity of the LPA over a range of wavelengths simultaneously. Furthermore, as the layers are assumed to be separated by a macro-scale distance within the farfield, the approach is inherently tolerant to layer misalignment. It is also highly amenable to current large-area lithography techniques, as each layer can be fabricated as an individual metasurface chip and then simply packaged together in the final device. As proof of concept, a metalens operating at 2000 and \SI{2340}{\nm} is designed that in simulation achieves focusing efficiencies of 65\% and 56\% with a NA of 0.11. The operating principle, design methodology and simulated lens performance will be discussed.

% \FloatBarrier
\section{Results and discussion}
\subsection{Device overview and working principle}
\begin{figure}[t]
\includegraphics{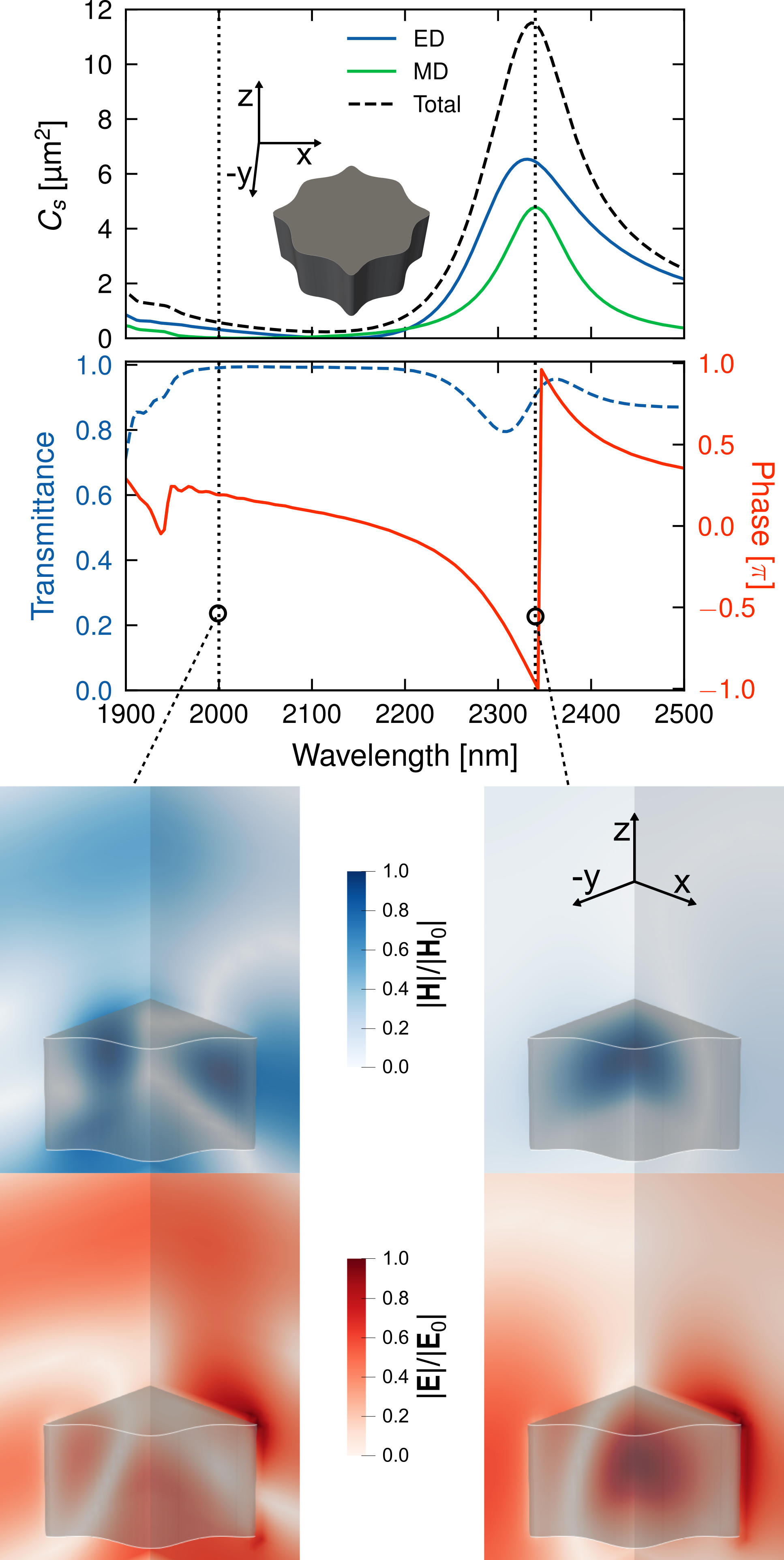}
\caption{\label{fig:work princ} Analysis of a Huygens' resonant meta-atom to illustrate device working principle. The example meta-atom is selected from the inverse designed \SI{2340}{\nm} library (inset) and is illuminated by an x-polarized planewave, normally incident on the substrate. A plot of the multipole-decomposition of the individual meta-atom scattering cross-section (top), and phase and transmittance through a periodic array (bottom) are shown. At the operating wavelengths (black dashed lines), plots of the normalized electric (E) and magnetic (H) fields through the yz and xz planes are provided.}
\end{figure}

In Huygens' metasurfaces the meta-atom array supports local spectrally overlapping electric (ED) and magnetic (MD) dipolar resonances which interfere to produce directional forward scattering and 2$\pi$ phase coverage \cite{staudeTailoringDirectionalScattering2013,deckerHighEfficiencyDielectricHuygens2015}. The archetypal meta-atom consists of a low aspect-ratio disc, where the spectral position of the resonances and thus the phase of the transmitted field is controlled by varying the disc radius. To achieve multiwavelength functionality within this regime, we leverage two characteristics of this interaction. The first is that within the immediate spectral vicinity of the resonance, the scattering cross-section is low and the transmittance high. This is because the dipolar resonances are of the lowest order, such that no other Mie-resonances are supported in the structure within some local bandwidth. This is illustrated in Figure~\ref{fig:work princ} in which the multipole decomposition, of an example meta-atom is shown, along with the field profiles at the operating wavelengths. The second is that away from resonance, the field is not strongly localized within the meta-atom (see inset field profiles in Figure~\ref{fig:work princ}) and does not strongly feel its geometry. Here, the metasurface behaves more like a thin-film with a small linear phase dispersion. Taken together, these traits enable Huygens' metasurfaces which simultaneously provide 2$\pi$ phase coverage at the resonance wavelength, while restricting the phase disturbance away from resonance. This is done primarily by decoupling the meta-atom array fill-factor from the spectral position of the resonances. Under this condition, it is possible for the phase accumulated off-resonance (dominated by a thin-film like interaction) to remain approximately constant over a meta-atom library. Unfortunately, the well known Huygens' disc geometry is not ideal in this instance as it requires the array fill-factor to be proportional to the applied resonant phase due to the only geometric degree of freedom being the disc radius. To overcome this, we generate meta-atom libraries using an inverse design scheme which enables many more geometric degrees of freedom while maintaining $C_4$ rotational symmetry and polarization insensitive operation. Using the obtained libraries, each layer in the multilayer metalens then applies a lensing phase profile with a focal length $F$ to the wavelength $\lambda$, given by 
\begin{equation}\label{eq:lens phase}
    \phi(r,F) = -\frac{2\pi}{\lambda}\sqrt{F^2+r^2} + C(\lambda),
\end{equation}
where $r$ is the radial distance from the lens center, and $C(\lambda)$ represents a wavelength dependent constant that contains the phase accumulated through the alternate layers. A schematic and render of the final multilayer device focusing two wavelengths is given in Figure~\ref{fig:schem and render}.
% \begin{figure}[t]
% \includegraphics{working_principle.pdf}
% \caption{\label{fig:work princ} Analysis of a Huygens' resonant meta-atom to illustrate device working principle. The example meta-atom is selected from the inverse designed \SI{2340}{\nm} library (see Figure~\ref{fig:meta-atom lib perf}a) and is illuminated by an x-polarized planewave. A plot of the multipole-decomposition of the individual meta-atom scattering cross-section (top), and phase and transmittance through a periodic array (bottom) are shown. Inset at the operating wavelengths (black dashed lines) are the electric and magnetic vector field profiles through the xz and yz planes respectively.}
% \end{figure}

\subsection{Inverse design of multiwavelength meta-atom libraries}
\begin{figure}[t]
\includegraphics{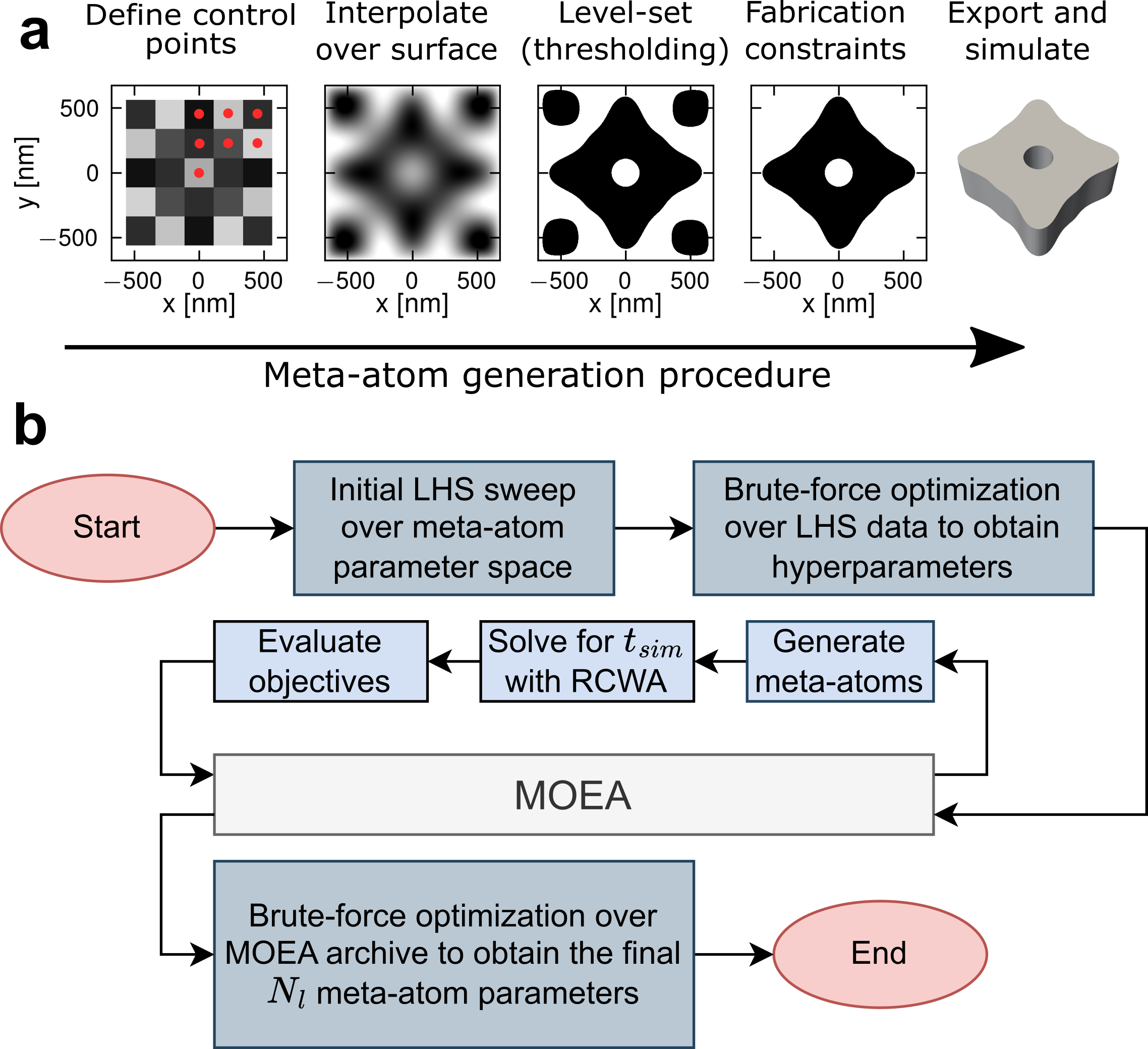}
\caption{\label{fig:inverse des proc} (\textbf{a}) Meta-atom parameterization and geometry generation procedure. The example given is for a $C_4$ symmetric geometry, which has 7 free parameters (red points) due to the $5\times 5$ grid of the control point array. These take on a value [0,1]. After interpolating, a level-set operation at a control surface value of 0.5 then defines the meta-atom boundary. (\textbf{b}) Flowchart of multiwavelength meta-atom library inverse design procedure.}
\end{figure}

To generate the multiwavelength libraries, we use a shape optimization inverse design scheme within the LPA, which is based off the work by Whiting et al. in Ref. \cite{whitingMetaatomLibraryGeneration2020}. The core of this scheme is an efficient meta-atom parameterization method (see Figure~\ref{fig:inverse des proc}a) in which the meta-atom geometry is obtained by interpolating an array of control points over the period to form a control surface. A level-set operation on this surface is then used to define the meta-atom boundary. This is a highly flexible method that allows arbitrary symmetries, and is ideal for creating polarization insensitive meta-atoms. Further fabrication constraints are imposed after obtaining the geometry through morphological or other algorithmic operations. The meta-atoms are simulated using Rigorous Coupled Wave Analysis (RCWA) at the wavelengths of interest. The complex transmittance ($t_{\text{sim}}$) obtained from RCWA is then used to construct a multiobjective function quantifying the meta-atom performance (for further details, see Supplementary S-III). The constructed objectives consider the desired multiwavelength performance both at the operating wavelengths and within $\pm\SI{10}{\nm}$ to avoid strong frequency dependence in the final designs, as this increases sensitivity to fabrication defects. The multiwavelength libraries are then generated by using the meta-atom control point array as optimization parameters within a gradient free multiobjective evolutionary algorithm (MOEA) \cite{hadkaBorgAutoAdaptiveManyObjective2013}. This allows for a whole $2\pi$ spanning library to be produced within a single optimization run containing roughly, 8000 objective evaluations. Further explanation of how a multiobjective optimization enables this process is available in Supplementary text. To speed up the MOEA optimization step, we first define meta-atom hyperparameters (the array period, height and non-resonant wavelength target phase) via a simple grid search based optimization using Latin Hypercube Sampling (LHS) \cite{bischlHyperparameterOptimizationFoundations2023} to obtain the initial dataset. The final libraries are then generated using an additional grid search over the whole MOEA archive. This is because at the end of the MOEA optimization, there are solutions spanning the entire modulation wavelength phase space, and only a subset of these that corresponds to the $N_l$ target modulation library phase levels are needed. Details on the benefits of selecting this discrete number for modulation phase levels are given in Section \ref{sec:meta-atom lib perf} and within the Supplementary text. The algorithm flowchart of this process is shown in Figure~\ref{fig:inverse des proc}b. Together with the first LHS step, a library is generated in roughly 10,000 total objective evaluations. As here we use the LPA, the simulation domain only consists of a single meta-atom unitcell. Combined with an efficient RCWA solver, this enables the whole optimization to be run on a workstation PC within one or two days.

\subsection{Meta-atom library performance}
\label{sec:meta-atom lib perf}
\begin{figure*}[t]
\includegraphics{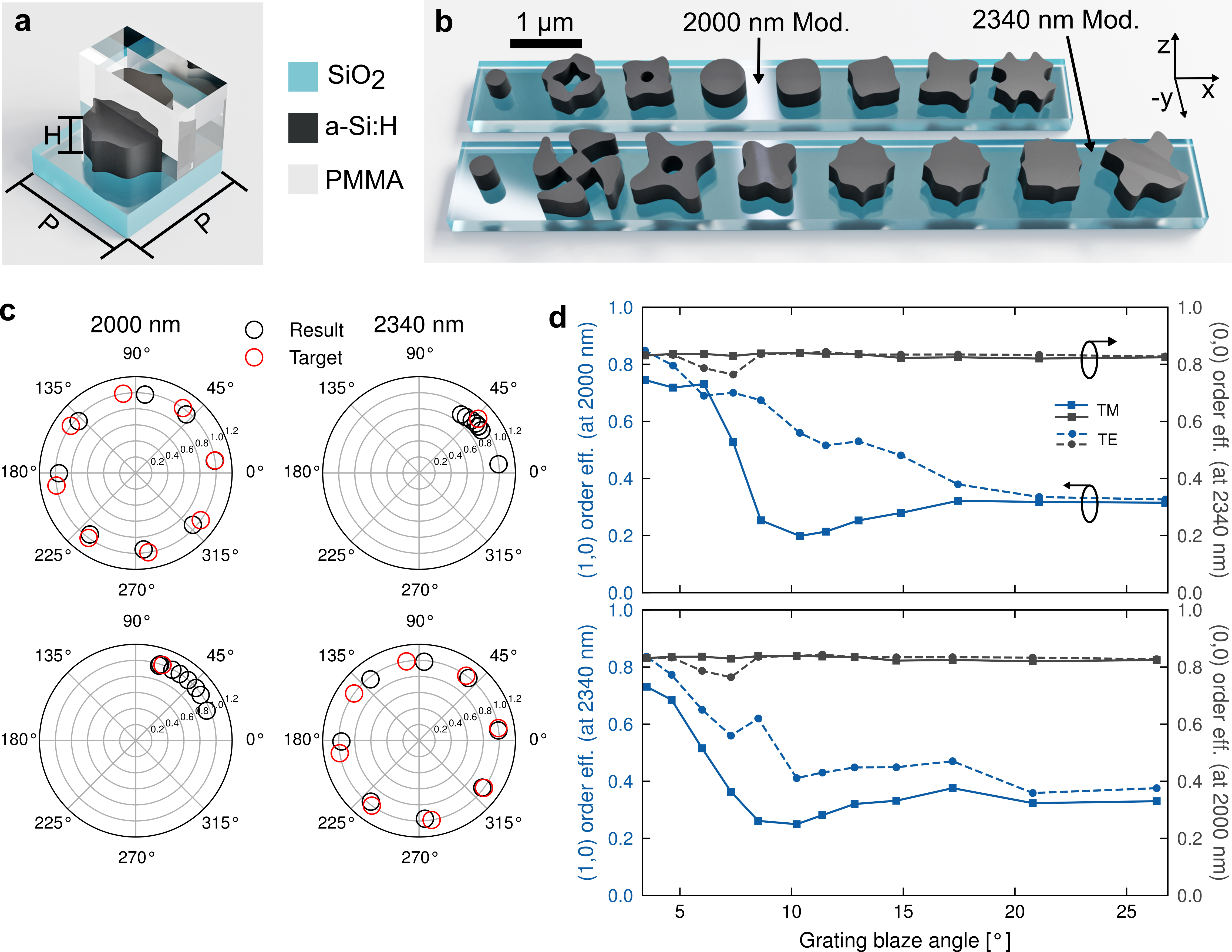}
\caption{\label{fig:meta-atom lib perf} (\textbf{a}) Example unitcell of a multiwavelength Huygens' metasurface. It consists of an a-Si meta-atom embedded in PMMA on a SiO$_2$ substrate. The period (P) of the unitcell and height (H) of the meta-atom is marked. (\textbf{b}) 3D render of finalized meta-atom libraries with the embedding PMMA not shown. For the \SI{2000}{\nm} library P = \SI{1110}{\nm}, H = \SI{320}{\nm}, and for the \SI{2340}{\nm} library: P = \SI{1320}{\nm}, H = \SI{380}{\nm}. (\textbf{c}) Polar plots of the complex meta-atom library transmission values calculated from RCWA. The finalized libraries are compared to an ideal target response (unitary transmission, constant phase spacing) as used in the inverse design procedure. The top row shows the \SI{2000}{\nm} library results and the bottom the \SI{2340}{\nm} library. (\textbf{d}) FDTD simulated performance of the libraries as metagratings designed to blaze into the (1,0) order at the modulation wavelength. The efficiency at both the (1,0) order for the library modulation wavelength and the (0,0) order for the alternate wavelength is shown. The light is normally incident from the substrate. The top panel shows the results for the \SI{2000}{\nm} library and the bottom the \SI{2340}{\nm} library. }
\end{figure*}
\begin{figure*}[t]
\includegraphics{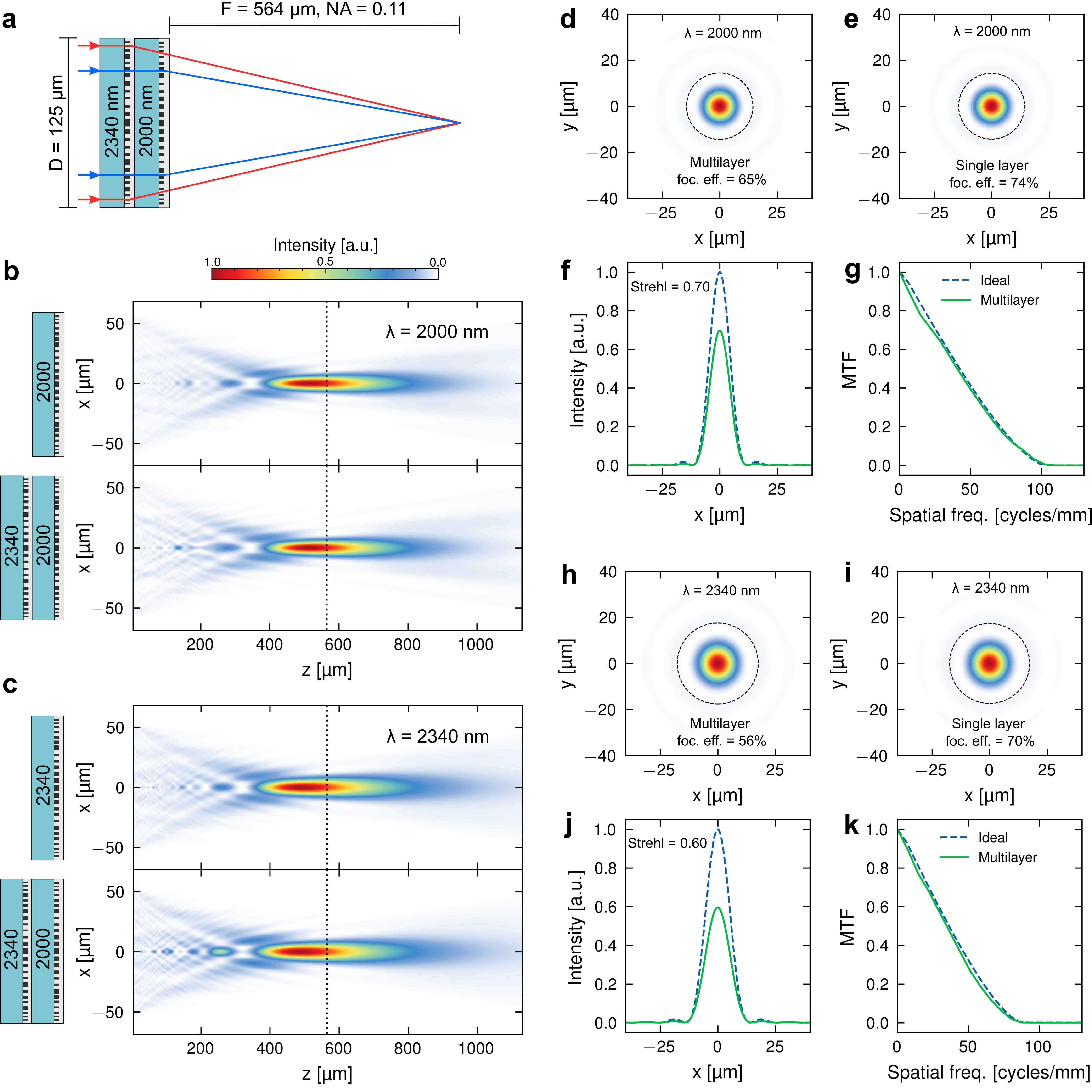}
\caption{\label{fig:whole lens perf} Results of multilayer metalens FDTD simulations. (\textbf{a}) Diagram of the simulated metalens. (\textbf{b}) Calculated farfield intensity cross-section along the propagation axis at \SI{2000}{\nm} with the design focal length indicated by the black dashed line. The top panel shows the field as propagated through only the single \SI{2000}{\nm} modulation layer, and the bottom panel the field when propagated through the whole multilayer metalens. (\textbf{c}) The same comparison at \SI{2340}{\nm}. (\textbf{d,e}) PSF's at the focal plane for both a single (\textbf{e}) or full multilayer lens (\textbf{d}) at \SI{2000}{\nm}. The black dashed circle represents the aperture used to calculate the focusing efficiencies (included in figure). (\textbf{f,g}) Comparison between the multilayer metalens at \SI{2000}{\nm} and ideal lens 1D PSF (\textbf{f}) and MTF (\textbf{g}). (\textbf{h-k}) Same results as the above four panels for the \SI{2340}{\nm} field.}
\end{figure*}
The designed meta-atom unitcells consist of a structured hydrogenated amorphous silicon (a-Si:H) layer on a fused silica (SiO$_2$) substrate embedded in a polymethyl methacrylate (PMMA) cladding (Figure~\ref{fig:meta-atom lib perf}a). The inverse design scheme is used to generate meta-atom libraries that structure the resonant wavelength with $N_l = 8$ phase levels. Using discrete rather than continuous phase levels at first seems counter-intuitive, as the geometry incurs rapid transitions over the surface that may cause the LPA to become invalid. However, we find in numerical studies that firstly the performance loss solely due to phase level discretization (known as phase quantization error in diffractive optics \cite{soiferComputerDesignDiffractive2013}) is less than 5\% (see discussion in S-II and Figure S1b), a result which is consistent even to high NAs as 8 phase levels is far from the Nyquist condition and well samples any periodic [0,$2\pi$] phase function. Secondly, in numerical trials of a Huygens' disc meta-atom geometry the difference between continuous phase levels and an 8 level library is negligible when analyzed as a blazed metagrating (Figure S2c), and is not expected to make a significant impact on the final metalens. The benefit of this approach is that it improves the off-resonant performance. Firstly, the complexity of optimizing a smaller subset of geometries for multiwavelength performance is reduced, which enables the phase deviation off-resonance to be lower. Secondly, the discrete number of meta-atom geometries minimizes the potential for off-resonance spatial phase gradients to occur, which would lead to unwanted deflection. To scale renders of the final libraries to modulate 2000 and \SI{2340}{\nm} are provided in Figure~\ref{fig:meta-atom lib perf}b. Although the geometries have complex shapes, the fabrication constraints applied during the inverse design loop restricts the minimum feature size to be larger than \SI{100}{\nm}. Performing multipole decomposition on the meta-atoms (see Figure S4), we find that the inverse design algorithm maintains the Huygens' resonance condition for all geometries, apart from the first two meta-atoms (counting from the left in Figure~\ref{fig:meta-atom lib perf}b) in each of the two libraries. This highlights the benefit of using an algorithmic design approach as although the Huygens' resonance provides phase coverage for the majority of the meta-atoms it is unsurprising that for low absolute phase (measured relative to the unitcell $t_\text{sim}$ absent the a-Si:H layer) a non-resonant geometry presents a better solution to the multiwavelength objective function. The polar plots in Figure~\ref{fig:meta-atom lib perf}c show the $t_\text{sim}$ values of the libraries compared to the target ideal library, which is used in the final step of the optimization procedure. The results show excellent on resonance phase coverage, with a small phase range for the off-resonant wavelengths and high transmittance for both cases. The libraries were also simulated as blazed metagratings using finite difference time domain (FDTD) to ascertain their multiwavelength beam deflection efficiency in an approximation of how they would perform locally within the metalens (Figure~\ref{fig:meta-atom lib perf}d). At the resonance wavelength, both libraries show high efficiency polarization-insensitive beam deflection into the (1,0) blaze order, for angles less than $\sim$\SI{7}{\degree}, with significant performance degradation as the angle increases. This poor performance at high deflection angles is a known limitation of Huygens' metasurfaces designed within the LPA and has been discussed in detail in Refs. \cite{arslanAngleselectiveAlldielectricHuygens2017, gigliFundamentalLimitationsHuygens2021}. Here we attribute it to unaccounted for cross-talk due to the broken periodicity in the metagrating, as well as the angular dispersion of the ED and MD Mie-modes \cite{arslanAngleselectiveAlldielectricHuygens2017} in addition to the emergence of other asymmetric modes at large in plane k-vectors which interfere with the Huygens' dipolar resonances \cite{gigliFundamentalLimitationsHuygens2021}. This limits the performance of high NA Huygens' metalenses designed using the LPA. Nevertheless, it should be noted that the 0.11 NA multiwavelength metalens presented in this work, has more than an order of magnitude higher NA than what would be achievable at centimeter scale diameters using standard single layer dispersion-engineering approaches that are limited by the restricted GD of current dielectrics (see further discussion in Supplementary S-I) \cite{presuttiFocusingBandwidthAchromatic2020, engelbergAchromaticFlatLens2021}. Furthermore, the metalens NA could be significantly increased under the same multilayer Huygens' metasurface paradigm via the use of a more sophisticated optimization scheme which goes beyond the simple LPA. It has been shown in Ref. \cite{arslanAngleselectiveAlldielectricHuygens2017} that the Huygens' resonance regime in nanodiscs can be implemented at larger deflection angles via judicious design of the disc aspect-ratio. This implies the angular bandwidth limitation is primarily a consequence of the LPA design method (in which only normally orientated 0th order S-parameters are considered) rather than the meta-atom platform itself. In addition, in Ref. \cite{caiInverseDesignMetasurfaces2020} the authors used a gradient free optimization scheme on hundreds of Mie-resonant nanodiscs to produce a 0.51 NA \SI{24}{\um} diameter cylindrical lens with a 60\% experimental focusing efficiency - thus demonstrating the potential of such a method. The trade-off being the increased computational overhead. Off-resonance, the multiwavelength libraries both have polarization-insensitive $\sim$83\% efficiency into the (0,0) order, which is independent of the blaze angle indicating the strong decoupling between the on and off-resonant wavefront shaping. The reduction from unitary (0,0) transmission is due to some scattering into higher orders, as the phase-restriction is not perfect.

\subsection{Whole metalens numerical simulation and analysis}
To characterize the multiwavelength metalens performance, the 8-level meta-atom libraries were used to generate two metasurfaces where each applies the hyperbolic phase in Eq. \ref{eq:lens phase} to its resonant wavelength determined by a simple phase look-up table. To remain within the optimal angular bandwidth of the meta-atom libraries, we use a lens phase profile with a NA of 0.11. The multilayer device was then simulated in FDTD. Due to limited computational resources, the metalens diameter was selected to be \SI{125}{\um}. A schematic of the simulated structure is provided in Figure~\ref{fig:whole lens perf}a. As the metalens layers are assumed to be spaced by a distance within the farfield, to avoid interlayer coupling effects whilst remaining within the computational budget, each layer was simulated individually with the field above the first layer used as the excitation source for the second layer. Further simulation details are available in the Supplementary text S-VII. In Figures \ref{fig:whole lens perf}b and c we compare the farfields of the full multilayer metalens to that of a single layer to demonstrate the effectiveness of this approach to decouple the on and off-resonant wavefront shaping. Both farfields demonstrate clear lensing with negligible change between their focal lengths, indicating the off-resonant response of the layers does not significantly disturb the applied resonant phase profile. The main distinction is the presence of increased intensity maxima closer to the lens. This results from a diffractive effect of the off-resonant layers that have a slight phase variation. They impose a weak radial periodicity over the lens's Fresnel zones, leading to scattering into higher orders. The point spread functions (PSFs) at the focal plane are plotted in Figures \ref{fig:whole lens perf}d,e,h and i for both the single and multilayer cases. These show a highly symmetric PSF, which due to the $C_4$ rotational symmetry of both the meta-atoms and full metalens array, is indicative of highly polarization-insensitive focusing. The shape of the PSF for both wavelengths is also essentially identical between the single and full multilayer device. The main difference is a reduction in focusing efficiency. This is defined using the commonly adopted approach as the ratio of optical power through an aperture with a diameter three times the PSF's full-width half maximum (FWHM) to the total power incident on the lens \cite{liInverseDesignEnables2022,shresthaBroadbandAchromaticDielectric2018,zhouMultilayerNoninteractingDielectric2018,arbabiMultiwavelengthPolarizationinsensitiveLenses2016} (dashed black circles in Figure~\ref{fig:whole lens perf}). This reduction in the efficiency is also attributed to the off-resonant diffraction. Nonetheless, the multilayer device still achieves high-efficiency polarization-insensitive focusing at 65\% and 56\% respectively, for the 2000 and \SI{2340}{\nm} incident fields. It should also be noted that due to the restricted metalens diameter of \SI{125}{\um}, diffraction limited focusing through an ideal lens at this NA has a maximum focusing efficiency of 86\% (see Supplementary text S-VI and Figure S6), such that the relative focusing efficiencies of the lens are 76\% and 65\%. Furthermore, larger centimeter-scale metalenses are expected to have improved performance as diffraction effects are reduced and the array size increases such that the LPA becomes more accurate for increasing lens area. In Figures \ref{fig:whole lens perf}f,g,j and k the one-dimensional (1D) PSF's along the x-axis are compared to that of an ideal lens, along with the 1D modulation transfer function (MTF). Here, the MTF is calculated from the Fourier transform of the PSF via \cite{goodmanIntroductionFourierOptics2005}:
\begin{equation}\label{eq:MTF def}
       \text{MTF} = \left|\frac{\int\int\text{PSF}(x,y)e^{i2\pi(f_x x + f_y y)}dxdy}{\int\int\text{PSF}(x,y)dxdy}\right|,
\end{equation}
where $f_x$ and $f_y$ are the spatial frequencies along the $x$ and $y$ axes. The ideal PSF used in the comparison is taken from the numerically propagated ideal lens farfield, as described in the Supplementary. These results show near diffraction-limited focusing performance, with the MTF matching almost exactly that of the ideal lens for both wavelengths. We find that the multilayer metalens in simulation produces close to ideal focusing performance, with the primary detraction simply being a reduction in the overall efficiency. 
\section{Conclusion}
We have presented a novel multiwavelength and polarization-insensitive metalens design which is suitable for centimeter scale lens diameters. In simulation, the design achieves a near diffraction-limited MTF with absolute focusing efficiencies  of 65\% and 56\% at operating wavelengths of \SI{2000}{\nm} and \SI{2340}{\nm} for a NA of 0.11. Compared to an ideal lens of the same dimensions and NA this is equivalent to a relative focusing efficiency of 76\% and 65\% respectively. The design consists of multiple Huygens' metasurface layers, where each layer structures the wavefront at the Huygens' resonance wavelength, while decoupling the wavefront shaping off-resonance. The layers are formed from meta-atom libraries with 8 phase levels, generated via an efficient shape-optimization inverse design method that maintains the $C_4$ rotational symmetry of the meta-atom geometry, thus ensuring polarization-insensitive operation. A key benefit of the approach is that the layers are designed to be separated by a distance within the farfield. This allows the multilayer design to be highly tolerant to layer misalignment. Additionally, it is well suited for current large-area nanofabrication techniques, as each layer can be fabricated individually and simply packaged together in the final device. In future work, the inverse design scheme presented here can be extended to consider multiple meta-atom interactions and larger operation angles, which can be used to further increase the efficiency and NA of this approach. Furthermore, the approach presented here is highly generalizable to any arbitrary multiwavelength phase profile beyond that of simple lensing. We anticipate that it will facilitate a variety of compact, large-area, multifunctional polarization-insensitive devices, such as color routers, vortex beams, double-helix PSFs, and other advanced phase mask engineering techniques. 
\section{Methods}
\small{
\subsection*{Shape-optimization inverse design}
The inverse design workflow was implemented in Python using standard scientific packages NumPy and SciPy. Fabrication constraint algorithms made use of the OpenCV module to access morphological operations. To generate 3D STL models for simulation in external software the open source software OpenSCAD was used and accessed via the SolidPython module. The MOEA algorithm used was the Python implementation of Borg \cite{hadkaBorgAutoAdaptiveManyObjective2013}. 
\subsection*{Numerical simulations}
The inverse design optimization made use of a RCWA solver implemented in Python based on the grcwa package presented in Ref. \cite{jinInverseDesignLightweight2020a}. Here, as we use a gradient free optimization, grcwa was modified to utilize an efficient JIT compiled backend via Numba. The RCWA simulations were run using 300 Fourier orders, which would take roughly \SI{1.5}{\second} for an evaluation at a single wavelength on a standard workstation PC. The complex permittivity for the RCWA simulations were extrapolated from experimental ellipsometric data (see Supplementary text S-VIII for more details).

The metagrating and large metalens simulations were performed using commercial FDTD software, Lumerical. The material optical properties were imported from the experimental ellipsometry data. For the metagrating simulations, the domain used Bloch-periodic boundary conditions in the x and y-directions with PML boundaries in the z-direction. They were excited with a normally incident plane wave source. Further details on the metalens simulations are available in the Supplementary text.
}
\section*{Acknowledgements}
The project on which this report is based was funded by the German Federal Ministry of Education and Research (BMBF) under the funding code 01QE2004B through the Eurostars programme (Project: E!113720 MuGasMeta) and the Australian Research Council (CE200100010). Responsibility for the content of this publication lies with the author. This work was also funded by the Deutsche Forschungsgemeinschaft (DFG, German Research Foundation) through the International Research Training Group (IRTG) 2675 “Meta-ACTIVE,” project number 437527638.

\bibliography{main.bib}% Produces the bibliography via BibTeX.

\end{document}

\preprint{}

%\title{Supplementary information for \\ ``Large area multiwavelength and polarization-insensitive metalenses via multilayer Huygens’ metasurfaces"}% Force line breaks with \\

\title{Supplementary information for \\ ``Large-area multiwavelength and polarization-insensitive metalenses via multilayer Huygens’ metasurfaces"}% Force line breaks with \\

\author{Joshua Jordaan}
    \affiliation{Institute of Solid State Physics, Friedrich Schiller University Jena, 07743 Jena, Germany}%Lines break automatically or can be forced with \\
    \affiliation{Institute of Applied Physics, Abbe Center of Photonics, Friedrich Schiller University Jena, 07745 Jena, Germany}
    \affiliation{ARC Centre of Excellence for Transformative Meta-Optical Systems (TMOS), Research School of Physics, Australian National University, Canberra ACT 2601, Australia}
 
\author{Alexander Minovich}%
    \affiliation{Institute of Solid State Physics, Friedrich Schiller University Jena, 07743 Jena, Germany}%Lines break automatically or can be forced with \\
    \affiliation{Institute of Applied Physics, Abbe Center of Photonics, Friedrich Schiller University Jena, 07745 Jena, Germany}

\author{Dragomir Neshev}
    \affiliation{ARC Centre of Excellence for Transformative Meta-Optical Systems (TMOS), Research School of Physics, Australian National University, Canberra ACT 2601, Australia}

% \author{Thomas Pertsch}
%     % \affiliation{Institute of Applied Physics, Abbe Center of Photonics, Friedrich Schiller University Jena, Albert-Einstein-Straße 15, Jena, 07745, Germany}%Lines break automatically or can be forced with \\
%     \affiliation{Institute of Applied Physics, Abbe Center of Photonics, Friedrich Schiller University Jena, 07745 Jena, Germany}
%     % \affiliation{Max Planck School of Photonics, Hans-Knöll-Straße 1, Jena, 07745, Germany}
%     \affiliation{Max Planck School of Photonics, 07745 Jena, Germany}
%     % \affiliation{Fraunhofer Institute for Applied Optics and Precision Engineering IOF, Albert-Einstein-Straße 7, Jena, 07745, Germany}
%     \affiliation{Fraunhofer Institute for Applied Optics and Precision Engineering IOF, 07745 Jena, Germany}

\author{Isabelle Staude}
    \affiliation{Institute of Solid State Physics, Friedrich Schiller University Jena, 07743 Jena, Germany}%Lines break automatically or can be forced with \\
    \affiliation{Institute of Applied Physics, Abbe Center of Photonics, Friedrich Schiller University Jena, 07745 Jena, Germany}
    
\date{\today}

\begin{abstract}
\end{abstract}

\maketitle

%\tableofcontents

\section{Limitations of dispersion-engineering for large-area multiwavelength metalenses}
An ideal lens in free space with a focal length $F$ will apply a frequency dependent phase profile given by:
\begin{eqnarray}\label{eq:id lens phase}
	\begin{split}
		% \phi(\omega,r) &= -\frac{2\pi}{\lambda}\left(\sqrt{F^2 + r^2} - F\right)\\
		% &= -\frac{\omega}{c}\left(\sqrt{F^2 + r^2} - F\right),
        \phi(\omega,r) &= -\frac{\omega}{c}\left(\sqrt{F^2 + r^2} - F\right),
	\end{split}
\end{eqnarray}
where $c$ is the speed of light in vacuum, $\omega$ frequency of the incident field and $r$ is the radial distance along the lens. The maximum group delay (GD) or linear phase-dispersion applied to the field is proportional to the lens radius $R$ and can be calculated via:
\begin{eqnarray}\label{eq:max GD}
	\begin{split}
		\text{GD}_{\text{max}} &= \left|\frac{\partial\phi(\omega,r)}{\partial\omega}\right|_{r=R} = \frac{\sqrt{F^2 + R^2} - F}{c}\\
		&\approx \frac{R^2}{2cF}\\
        &\approx \frac{\text{NA}\times R}{2c},
	\end{split}
\end{eqnarray}
where NA is the numerical aperture of the metalens. We can now clearly see that the limited GD values that locally-interacting nanostructures can provide introduces a severe trade-off between the NA and lens size with
\begin{equation}
    \text{NA}\times R \leq 2c\text{GD}_{\text{max}}.
\end{equation}
Current dielectrics and nanofabrication capabilities are in the best case able to provide maximum GD values at optical frequencies on the order of $\sim$\SI{100}{\femto\second} \cite{heidenDesignFrameworkPolarizationinsensitive2022,liMetaopticsAchievesRGBachromatic2021,shresthaBroadbandAchromaticDielectric2018}. At centimeter sized metalenses this limits the NA to less than $\sim$0.01. In reality however, this is a significant approximation which does not consider the frequency bandwidth that the linear GD response is maintained over, which further constrains the design trade-offs and the maximum NA achievable. More detailed analysis can be found in Refs. \cite{presuttiFocusingBandwidthAchromatic2020,engelbergAchromaticFlatLens2021}.

\section{Meta-atom library size considerations}
\label{sec:SI Meta-atom library size considerations}
\begin{figure}[t]
\includegraphics{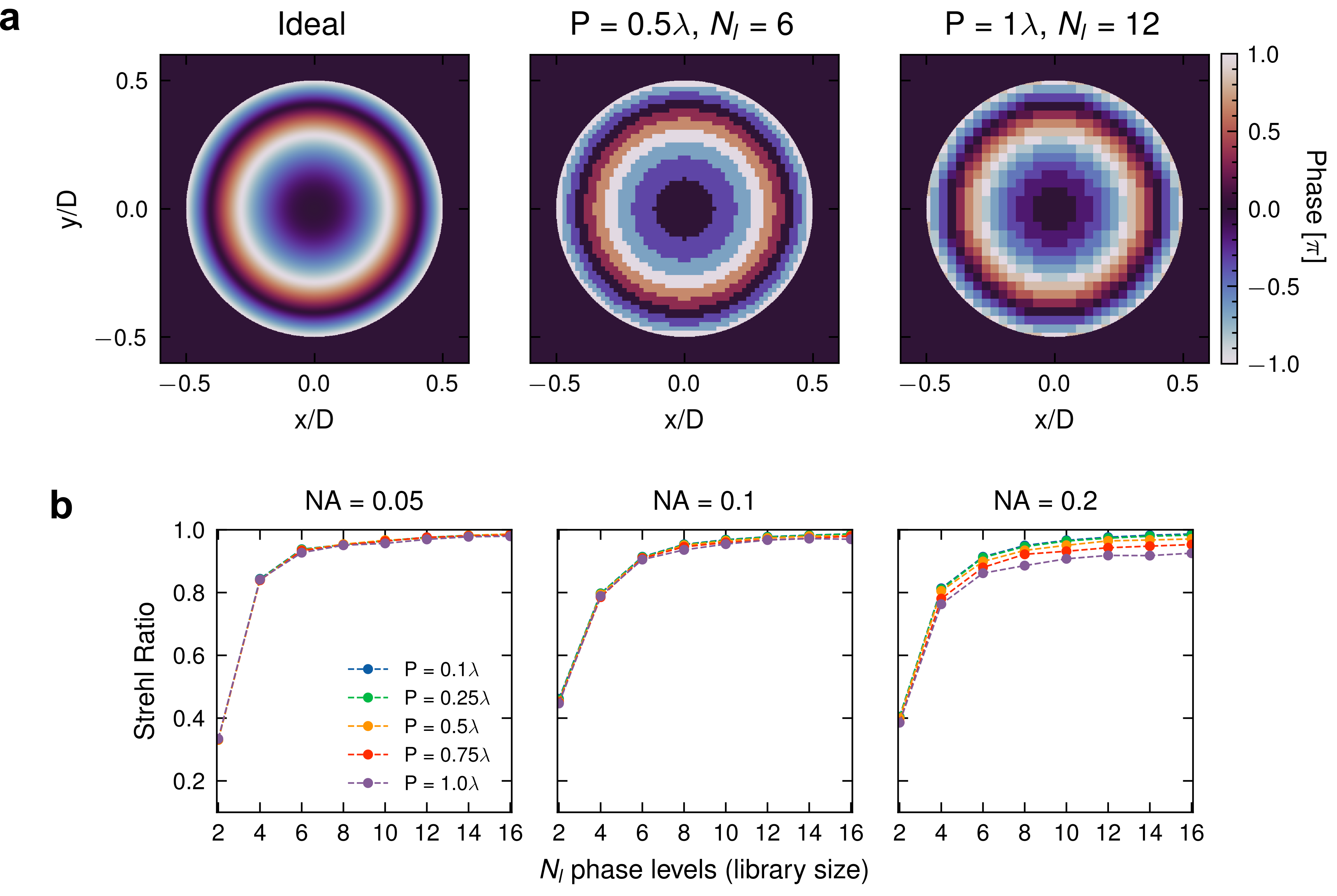}
\caption{\label{fig:toy metalens strehl plots} (\textbf{a}) Comparison of an Ideal continuous lensing phase profile, compared to that of different phase discretization (period size $P$) and quantization (number of discrete evenly spaced phase levels $N_l$) conditions. (\textbf{b}) Results of propagating the field through a toy metalens model with varying NA. The Strehl ratio is calculated for different discretization and quantization conditions to quantify the focussing performance.}
\end{figure}

When designing metalenses while using the locally periodic approximation (LPA) each meta-atom is associated with a single 0th order scattering response such that we approximate what is ideally a continuous phase profile by using meta-atoms of a finite period (phase discretization) and number of phase levels (phase quantization) (see Figure \ref{fig:toy metalens strehl plots}a). These effects are similar to those seen in diffractive optics \cite{soiferComputerDesignDiffractive2013}. The phase quantization and discretization act together to limit the maximum NA of a metalens and its efficiency. The NA is limited by the effect of the discretized sampling of the lens phase profile, leading to aliasing. The gradient of a hyperbolic lens phase increases towards an asymptote with the radial distance from the center. As such, for a given meta-atom period $P$ there exists a maximum NA that is sampled at the edge by a minimum phase interval of $2\pi/N_{\text{min}}$ over the surface, where $N_{\text{min}} = 2$ corresponds to the Nyquist limit, thus restricting the achievable NA. The relationship between these can be found by using the wavelength dependent form of Eq. \ref{eq:id lens phase}, and setting the phase gradient provided by two adjacent meta-atoms equal to the spatial gradient at the radius of the lens:
\begin{eqnarray}
    \begin{split}
		\frac{2\pi}{N_\text{min}P} &= \left|\frac{\partial\phi(\lambda,r)}{\partial r}\right|_{r=R} = \frac{2\pi R}{\lambda\sqrt{F^2+R^2}}\\
		&= \frac{2\pi}{\lambda}\text{NA}\\
        \Rightarrow N_\text{min} &= \frac{\lambda}{\text{NA}\times P},
	\end{split}
\end{eqnarray}
where $\lambda$ is the wavelength of the incident field. For a given NA and wavelength, the maximum period size that avoids aliasing is found using the above result and setting $N_{\text{min}} = 2$. Furthermore, for a given wavelength, increasing the NA without decreasing the period will result in lower $N_\text{min}$ and subsequent poorer phase sampling near the lens edge - which will tend to decrease the lens performance. This analysis also shows the principle of a large-area multiwavelength metalens in which each operating wavelength is structured individually. In this instance, the diameter of the lens no longer bounds the focusing phase profile, and for a given NA, the lens can be scaled to arbitrary dimensions. To quantify these effects on performance, we created a toy metalens system consisting of a $100\lambda$ wide circular aperture with unitary transmittance and quantized and discretized lens phase profile. The profile is quantized with $N_l$ evenly spaced phase levels, and the meta-atom period size sets the spatial discretization. The field through the toy metalens is propagated to the focal plane using a Python implementation of the band-limited angular spectrum method (BLAS) \cite{matsushimaBandlimitedAngularSpectrum2009}, and compared to that of an ideal (continuous phase profile) metalens to calculate the Strehl ratio. The results of which are in Figure \ref{fig:toy metalens strehl plots}b. Here we see the interdependency of the NA and meta-atom period size, with larger NA requiring smaller $P$ to reach near unitary Strehl ratio. Importantly, the effect of the quantization saturates quickly with increasing phase levels. Where in each case, $\sim$95\% of the maximum Strehl ratio is obtained for a given NA and $P$ at a value of $N_l = 8$. This is because 8 phase levels are already far from the Nyquist limit and well sample the phase profile, such that the effects of quantization become negligible. These results are in keeping with similar analysis typically performed for diffractive multilevel lenses \cite{soiferComputerDesignDiffractive2013}.

\begin{figure}[t]
\includegraphics{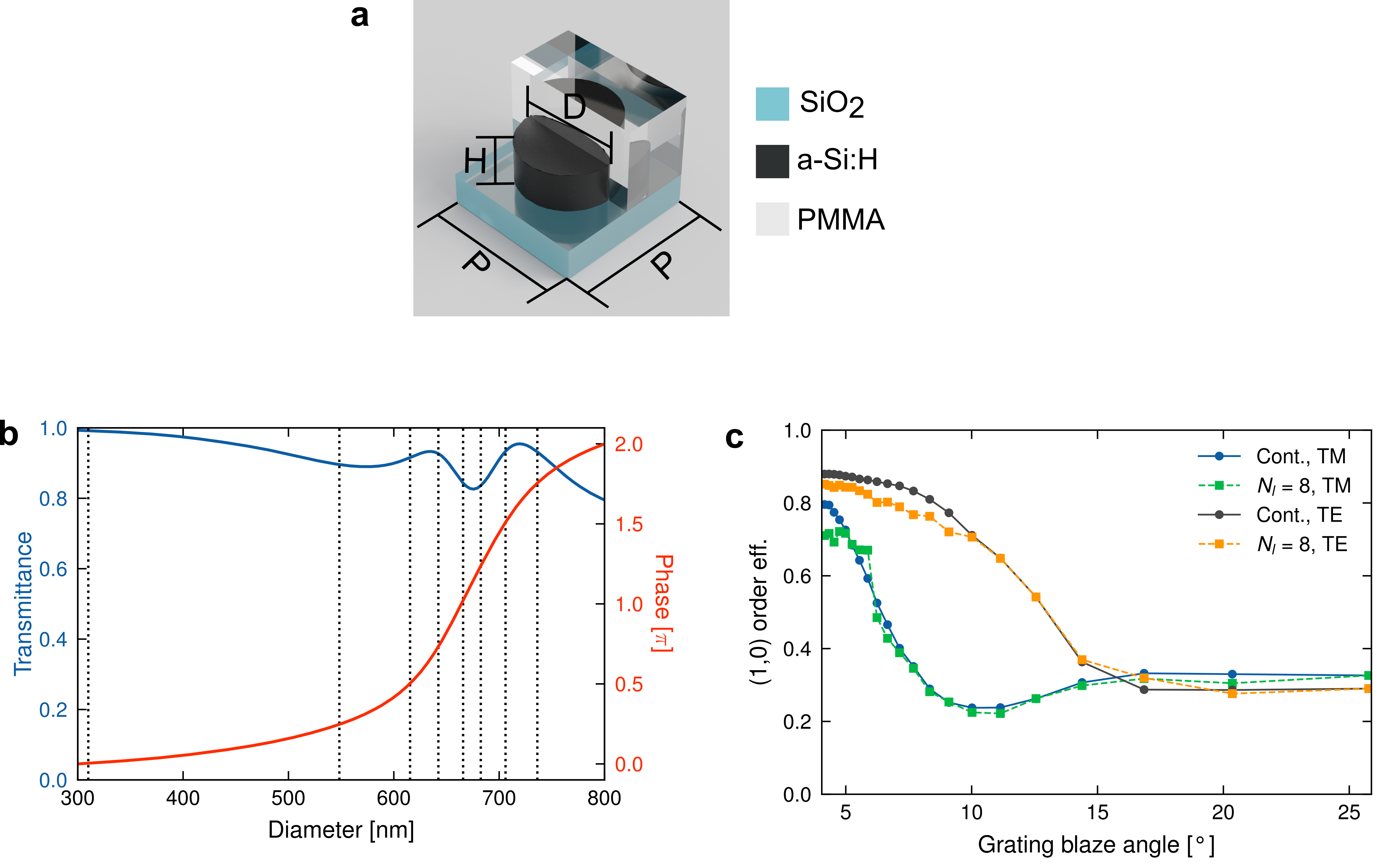}
\caption{\label{fig:cont. vs quant. library analysis} Analysis of a continuous vs discrete $N_l = 8$ phase level Huygens' meta-atom library. (\textbf{a}) Meta-atom unit-cell designed to be resonant at \SI{2000}{\nm}. The period size $P$ is \SI{1150}{\nm} and the a-Si:H meta-atom has a height $H$, of \SI{325}{\nm}. (\textbf{b}) Phase and transmittance of the meta-atom for varying diameters, calculated using RCWA. The black dashed lines indicated the diameters used in the discrete phase level library. (\textbf{c}) Results of analyzing both libraries as blazed metagratings for varying blaze angle, simulated in FDTD.}
\end{figure}

Another important factor concerning how the meta-atom library size (or phase quantization condition) impacts performance is dependent on the specific meta-atom platform itself. A meta-atom library with discrete rather than continuously varying geometries will incur more rapid spatial variations in the local optical properties and is therefore less likely to satisfy the LPA. This can lead to unaccounted for nearfield interactions and coupling between neighboring meta-atoms. To quantify this effect within a Huygens' resonant system, we study a low aspect ratio disc geometry that is designed to have a Huygens' resonance at \SI{2000}{\nm} (see Figure \ref{fig:cont. vs quant. library analysis}a). By varying the disc diameter, continuous $2\pi$ phase coverage and high transmittance is achieved (Figure \ref{fig:cont. vs quant. library analysis}b). We then select 8 disc diameters corresponding to evenly spaced phase levels (dashed black lines in Figure \ref{fig:cont. vs quant. library analysis}b), and compare the performance of the $N_l = 8$ phase level library with that of the continuously varying geometry. This is done by simulating blazed metagratings for varying blaze angle using FDTD, and recording the deflection efficiency into the blaze order for both TE and TM polarizations (results given in Figure \ref{fig:cont. vs quant. library analysis}c). For larger blaze angles ($\geq\SI{11}{\degree}$) the performance is practically identical between the continuous and discrete libraries. This is due to other effects inherent to Huygens' metasurfaces at larger angles beginning to dominate the deflection performance (discussed in main text and in Ref. \cite{gigliFundamentalLimitationsHuygens2021}), rather than the inherent difference in phase sampling. However, at low angles the discrete library only suffers at most a 10\% decrease in deflection efficiency. This is because at these low phase-gradients the discrete library metagrating is formed from larger sections of repeated elements, which tends to minimize the LPA discrepancy. Furthermore, although the Huygens' resonance is more prone to cross-talk than other non-resonant or waveguide type geometries - it is still fundamentally a local resonance and does not feature large lattice resonances which are sensitive to differences in geometry. 

\newpage
\section{Objective functions used in inverse design scheme}
We encode the desired multiwavelength meta-atom response into an unconstrained multiobjective function, such that the optimization problem has the form:
\begin{eqnarray}
	\begin{split}
		\text{Minimize: } &\boldsymbol{O}(\mathbf{x};\boldsymbol{\lambda}^\alpha,\boldsymbol{\lambda}^\beta,\beta,\mathbf{w}) = \left[ O_1,O_2\right] ^\text{T}\\
		\text{Subject to: } &0 \leq x_i \leq 1~\forall i\in\{1,2,...,n\},\\
	\end{split}
\end{eqnarray}
where $\mathbf{x}$ is the vector of control points that parameterizes the meta-atom geometry and is also the free variables within the optimization loop. The other terms in the function are hyperparameters which must be provided. Here, the objective function must take into account both on and off-resonant responses, and in addition we include the response $\pm\SI{10}{\nm}$ from the operation wavelengths to avoid strong frequency dependence in the final designs, as this increases sensitivity to fabrication defects. The additional terms are then $\boldsymbol{\lambda}^\alpha$ which is a vector containing the resonance wavelength and terms $\pm\SI{10}{\nm}$. Similarly, $\boldsymbol{\lambda}^\beta$ contains the wavelengths at and near the off-resonance wavelength, $\beta$ is the target off-resonance phase, and $\mathbf{w}$ is a vector of weights that are applied to sub-terms within the objectives to tune their behavior. The two objectives $O_1$ and $O_2$ are identical apart from a term that encapsulates the resonant phase response:
\begin{eqnarray}
	\begin{split}
		O_1 &= w_{\alpha}O_{\alpha} + w_{\beta}O_{\beta} + O_{\alpha\text{-phase}}\\
        O_2 &= w_{\alpha}O_{\alpha} + w_{\beta}O_{\beta} - O_{\alpha\text{-phase}},
	\end{split}
\end{eqnarray}
where subscript $\alpha$ and $\beta$ denote the weights and objective sub-functions to capture the performance on and off-resonace, and
\begin{equation} 
    O_{\alpha\text{-phase}} = \frac{\text{angle}(t_{\text{sim}}(\mathbf{x},\lambda_\alpha))}{2\pi},
\end{equation}
with $\lambda_\alpha$ equal to the resonance wavelength and $t_{\text{sim}}$ returning the complex transmission $S_{21}$ parameter at $\lambda^\alpha$ from the RCWA meta-atom simulation. Using this structure for the objective space defines a Pareto-front which satisfies $O_1 + O_2 = w_{\alpha}O_{\alpha} + w_{\beta}O_{\beta}$, such that every unique value of $O_{\alpha\text{-phase}}$ determines a new non-dominated solution for a given $(w_{\alpha}O_{\alpha} + w_{\beta}O_{\beta})$. This feature of multiobjective optimization enables the inverse scheme to generate a library spanning the full $2\pi$ phase space on-resonance within a single run. More details on this process are provided in Ref. \cite{whitingMetaatomLibraryGeneration2020}. The $O_\alpha$ and $O_\beta$ terms then capture the meta-atom response, which should be globally minimized for all points in the $2\pi$ resonant phase space. We define them as:
\begin{eqnarray}
	\begin{split}
		O_\alpha &= w_{\text{trans}}[1 - \text{mean}(|t_{\text{sim}}(\mathbf{x},\boldsymbol{\lambda}^\alpha)|^2)] + w_{\text{phase}}\frac{\text{range}\left(\text{angle}(t_{\text{sim}}(\mathbf{x},\boldsymbol{\lambda}^\alpha))\right)}{2\pi}\\
        O_\beta &= w_{\text{trans}}[1 - \text{mean}(|t_{\text{sim}}(\mathbf{x},\boldsymbol{\lambda}^\beta)|^2)] + w_{\text{phase}}\frac{\left|\text{angle}(t_{\text{sim}}(\mathbf{x},\lambda_\beta)e^{-i\beta})\right|}{2\pi},
	\end{split}
\end{eqnarray}
with $\lambda_\beta$ equal to the off-resonance operating wavelength. The above objectives are minimized when the meta-atom response has a high transmittance over the $\boldsymbol{\lambda}^\alpha$ and $\boldsymbol{\lambda}^\beta$ ranges, has a flat phase-response over $\boldsymbol{\lambda}^\alpha$ and has a phase value at $\lambda_\beta$ equal to $\beta$. The weights applied to the objectives are determined empirically. We find reasonable performance for values of $[w_{\alpha},w_{\beta},w_{\text{trans}},w_{\text{phase}}] = [1,0.5,0.6,0.4]$.

\newpage

\section{Meta-atom library geometry and array fill-factor}
\begin{figure}[ht]
\includegraphics{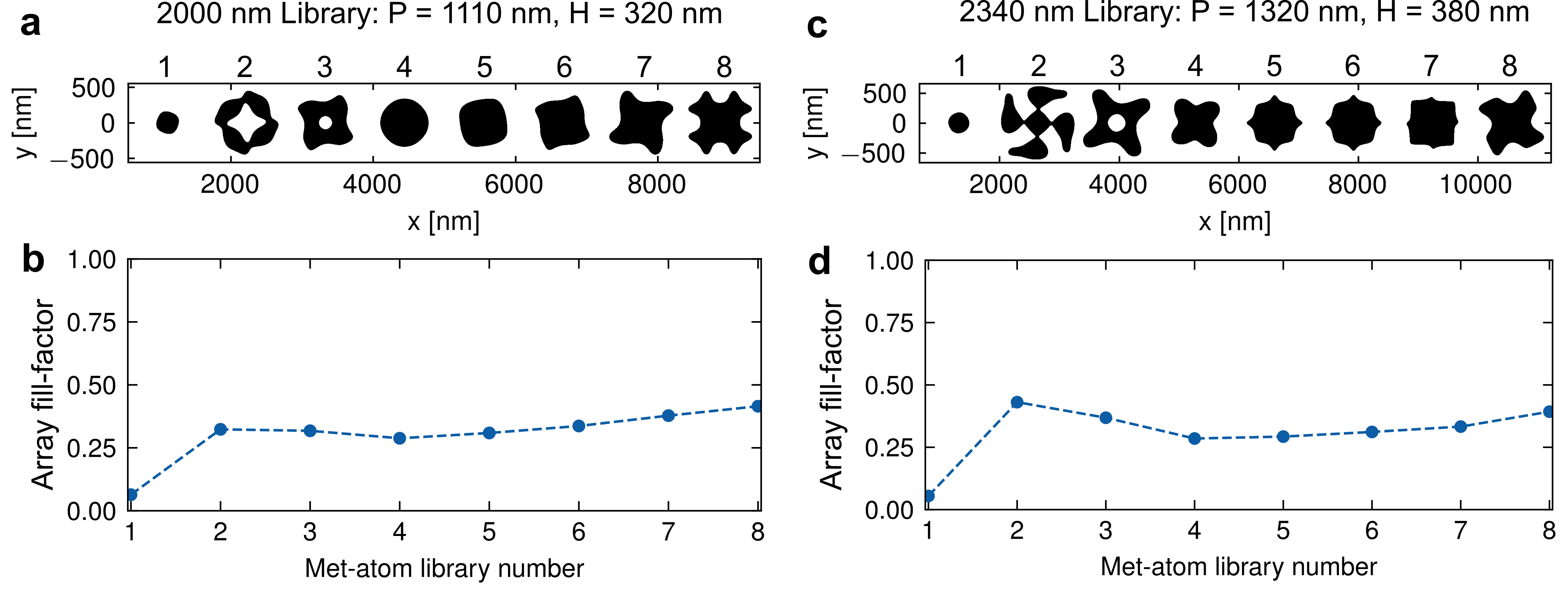}
\caption{\label{fig:ma geom fill frac} Meta-atom cross-sections and library numbering scheme for the \SI{2000}{\nm} (\textbf{a}) and \SI{2340}{\nm} (\textbf{c}) resonant libraries. The array fill-factor is calculated for each of the meta-atoms in both libraries in   (\textbf{b}) and (\textbf{d}).}
\end{figure}

Cross-section plots of the finalized meta-atom geometries are provided in Figures \ref{fig:ma geom fill frac}a and c. These also indicate the meta-atom library phase-level numbering convention. In Figures \ref{fig:ma geom fill frac}b and d we calculate the array fill-factor of the meta-atom geometry over each of the libraries. This is defined as: $(\text{meta-atom cross-section area})/P^2$. From these plots, we see that the fill-factor is non-monotonic over the library, which serves to minimize off-resonant deflections. It is also kept relatively constant, apart from the first phase level, in both libraries. 

\section{Multipole decomposition of meta-atom libraries}
\begin{figure}[ht]
\includegraphics{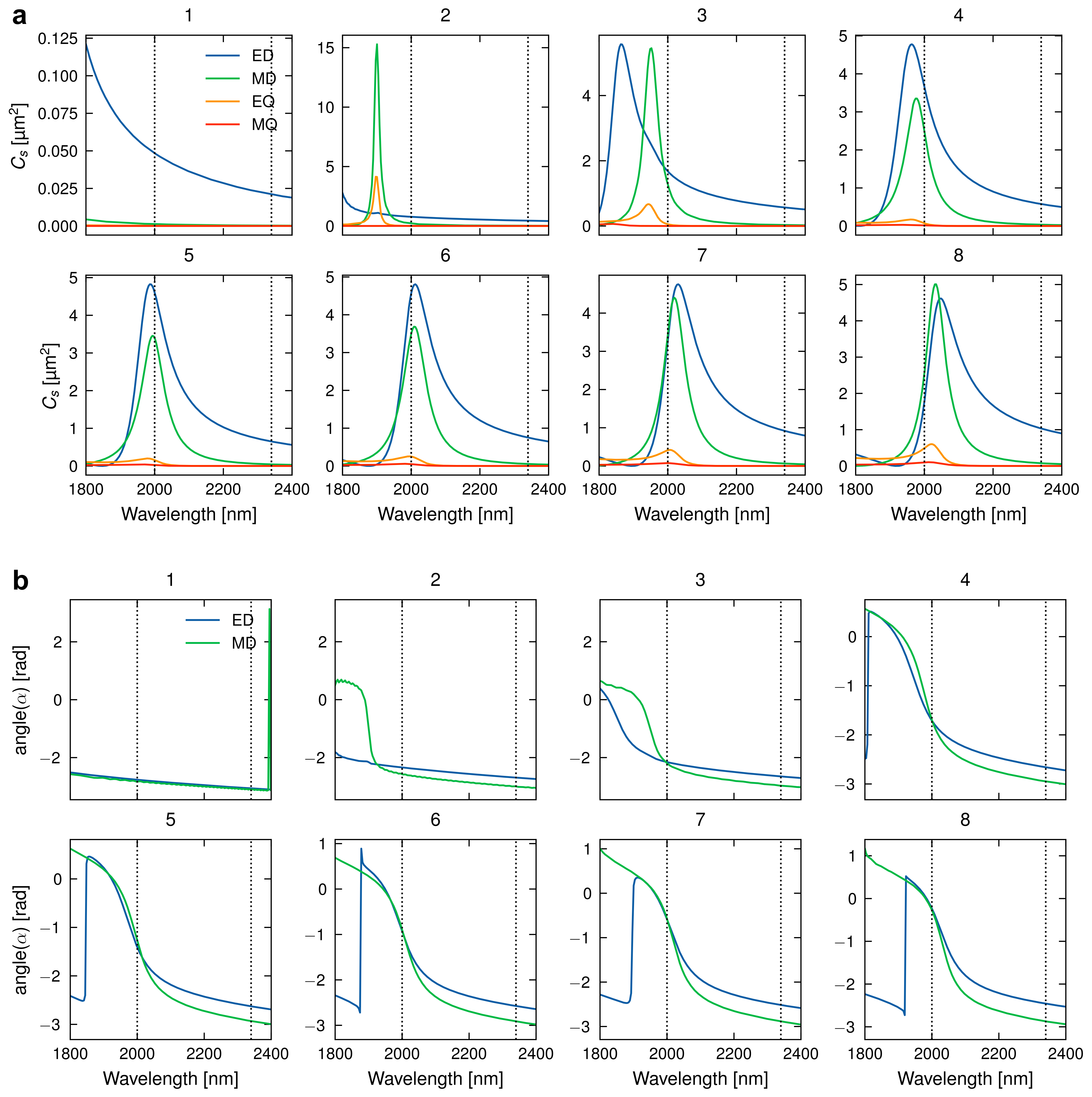}
\caption{\label{fig:multipole decomp 2000} Multipole decomposition of the \SI{2000}{\nm} resonant library. (\textbf{a}) Scattering cross-sections of the multipole coefficients up to quadrapolar terms. (\textbf{b}) Phase of the complex electric dipole (ED) and magnetic dipole (MD) multipole terms. In all plots, the dashed black lines correspond to the two operating wavelengths.}
\end{figure}

\begin{figure}[ht]
\includegraphics{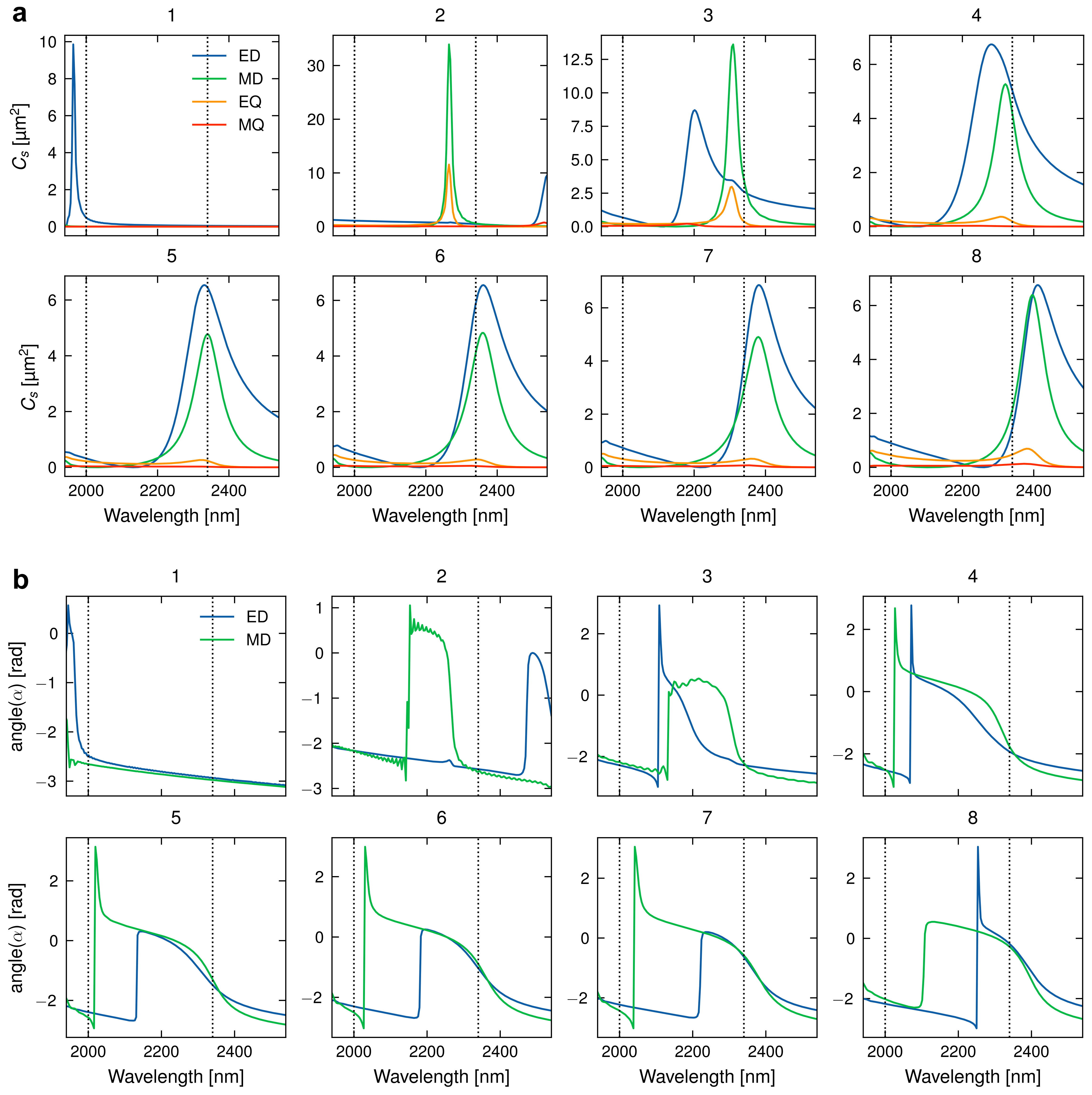}
\caption{\label{fig:multipole decomp 2340} Multipole decomposition of the \SI{2340}{\nm} resonant library. (\textbf{a}) Scattering cross-sections of the multipole coefficients up to quadrapolar terms. (\textbf{b}) Phase of the complex electric dipole (ED) and magnetic dipole (MD) multipole terms. In all plots, the dashed black lines correspond to the two operating wavelengths.}
\end{figure}

Multipole decompositions were performed on both meta-atom libraries. This was done by simulating the individual meta-atoms in an identical array using FDTD via the software package Lumerical, and then extracting the complex multipole coefficients using a custom Python implementation of the  multipole decomposition theory presented in Ref. \cite{grahnElectromagneticMultipoleTheory2012}. The results of which are provided in Figures\ref{fig:multipole decomp 2000} and \ref{fig:multipole decomp 2340} for the \SI{2000}{\nm} and \SI{2340}{\nm} resonant libraries, respectively. Here we see that apart from the first two phase levels of each library (which are not Huygens' resonant), the magnetic (MD) and electric (ED) dipoles terms radiate in-phase at the resonance wavelength. They also have dominant and similar magnitude ED and MD scattering cross-section contributions. These two features define the Huygens' resonant regime. Furthermore, we find that the phase of the ED and MD coefficient varies monotonically through the library, which reveals how the inverse-design scheme generates $2\pi$ phase coverage.

\newpage
\section{Ideal lens performance}
\begin{figure}[ht]
\includegraphics{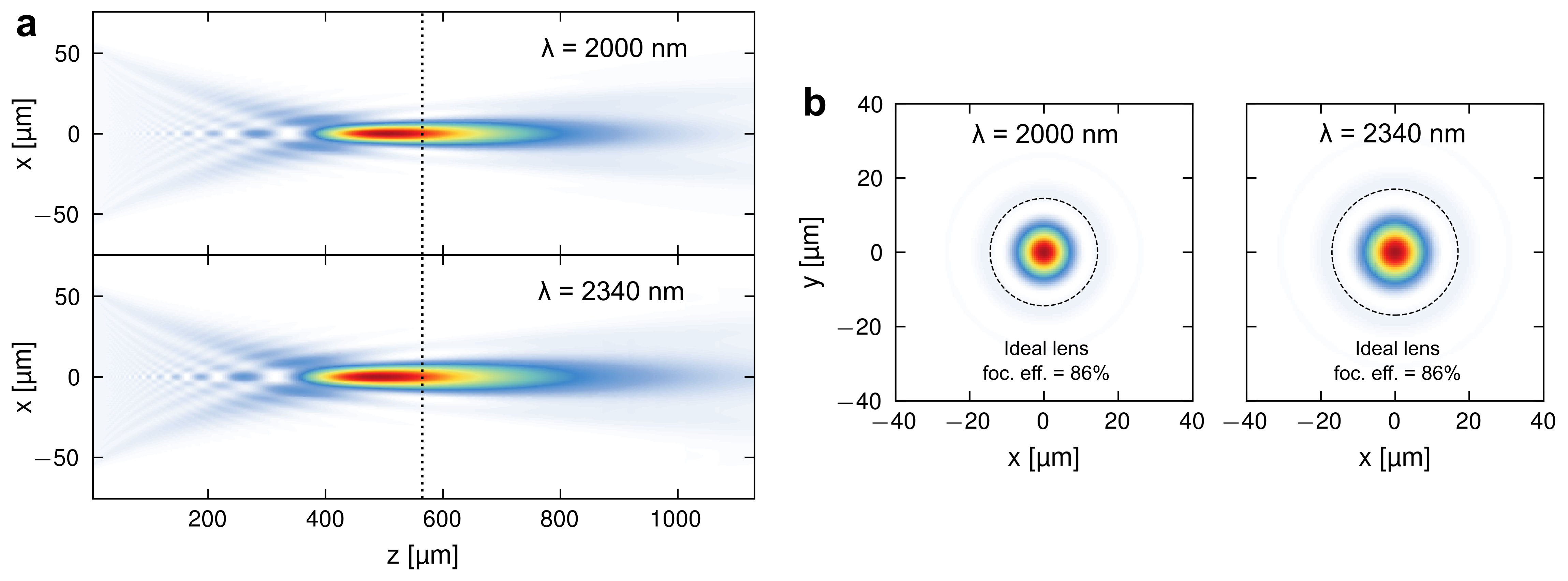}
\caption{\label{fig:ideal lens perf} Intensity cross-sections (\textbf{a}) and PSF's at the focal plane (\textbf{b}) of an ideal lens operating at each of the design wavelengths. The lens diameter and NA is the same that used in multiwavelength metalens simulations from the main text. The dashed lines in (\textbf{b}) indicate the designed focal length. The dashed black circles in (\textbf{b}) indicate the area through which the focussing efficiency is calculated, it has a diameter equal to $3 \times \text{FWHM}$ of the PSF.}
\end{figure}
Mathematically ideal lenses of the same dimension and NA of the multilayer multiwavelength metalens were simulated for each operating wavelength. This consisted of a \SI{125}{\um} diameter circular aperture with unitary transmittance and an ideal hyperbolic lensing phase function (Eq. \ref{eq:id lens phase}). The fields through this aperture were propagated to the focal plane via the BLAS method (as described in S-II), and the focussing efficiency calculated. The results in Figure \ref{fig:ideal lens perf} show that for this NA and diameter, the effects of diffraction still significantly impact the ideal lens performance, with a large proportion of the optical power in the outer lobes of the PSF. This reduces the calculated ideal focussing efficiency to 86\%. Furthermore, there is a shift in the intensity maxima to a shorter distance than the designed focal length. Both of these effects are also seen in the farfields of the simulated metalenses, which indicates a likely increase in performance for larger metalens apertures, where diffraction is less dominant.

\section{Additional simulation details and field profiles over multilayer metalens}
\begin{figure}[ht]
\includegraphics{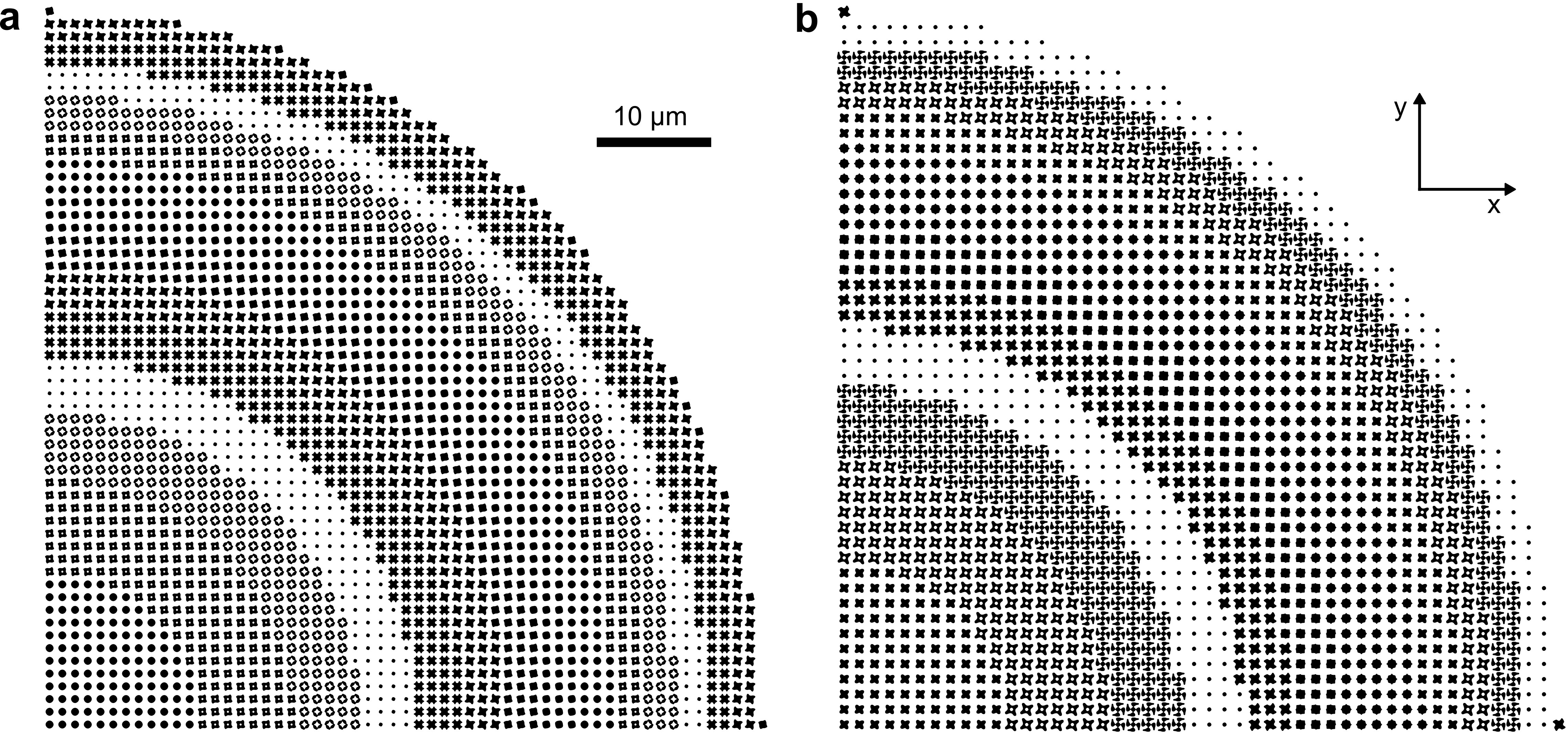}
\caption{\label{fig:meta-atom arrays} Full simulated meta-atom arrays. In (\textbf{a}) the \SI{2000}{\nm} resonant layer is depicted and (\textbf{b}) shows the \SI{2340}{\nm} layer.}
\end{figure}

\begin{figure}[ht]
\includegraphics{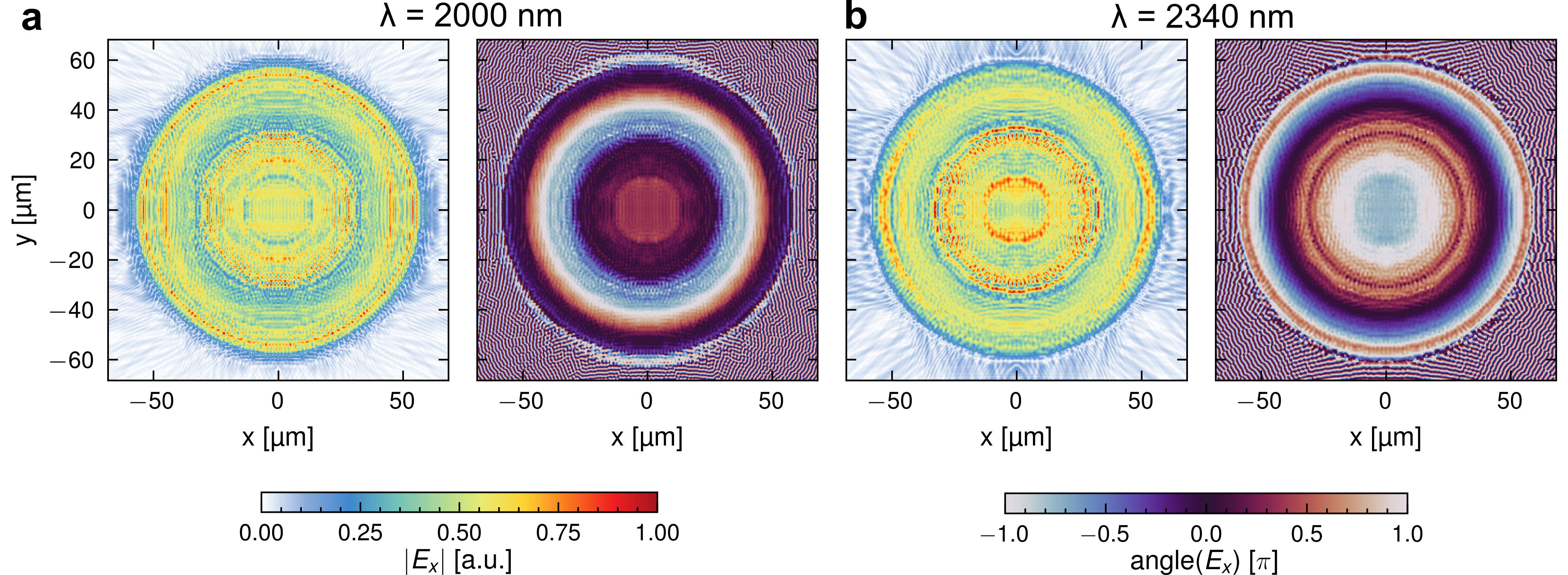}
\caption{\label{fig:field profiles} Phase and amplitude of the x-polarized electric field above the final metalens layer, which is propagated to the focal plane, to determine the PSF. In (\textbf{a}) the field \SI{2000}{\nm} is given and in (\textbf{b}) the \SI{2340}{\nm} field.}
\end{figure}

The multilayer metalens was simulated using the FDTD software Lumerical. To avoid interlayer coupling effects whilst remaining within a limited computational budget, the layers are simulated individually. The full stack is then simulated by exporting the field a distance one free-space wavelength above the first metasurface layer, and then importing this field as the source for the second layer. This method avoids evanescent field components coupling between layers, and closely approximates two layers separated by a distance within the farfield. As the meta-atoms have $C_4$ rotational symmetry and are arranged in a square array, the full metalens itself is also $C_4$ symmetric. This symmetry is utilized to reduce the computational domain to $1/4$ of the full metalens array. As we excite the lens with a field polarized along the x-axis, for a lens centered on the origin this entails using a perfect electrical conducting (PEC) boundary along the y-axis and perfect magnetic conducting (PMC) boundary along the x-axis. For all other boundaries, we use the perfectly matched layer (PML) boundary to simulate open space. We excite the first layer with a custom source, which consists of a circular region of unitary amplitude x-polarized electric field with the same diameter as the metalens. To improve the simulation convergence, the custom source has a \SI{5}{\um} wide Gaussian apodization applied to the edge of the radius. This is to avoid exciting the non-physical high spatial frequencies associated with field discontinuities where the source vanishes between adjacent mesh nodes. In Figure \ref{fig:meta-atom arrays} the full meta-atom arrays that were simulated for each layer are shown to scale. In Figure \ref{fig:field profiles} the phase and amplitude of the field directly above the multilayer metalens that was used to propagate to the focal plane is also provided.

\newpage
\section{Material permittivity data}
\begin{figure}[ht]
\includegraphics{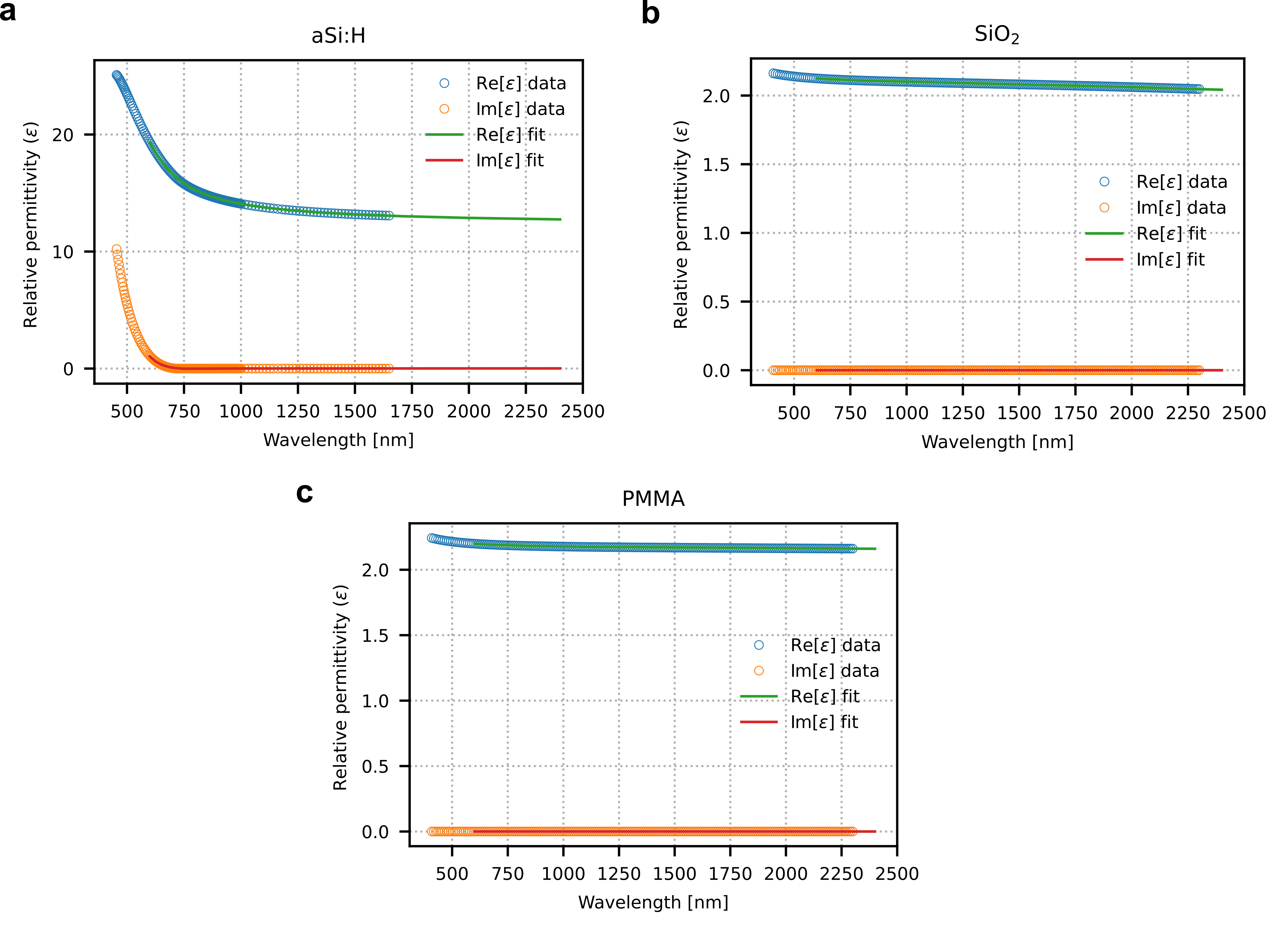}
\caption{\label{fig:material data} (\textbf{a} - \textbf{c}) Ellipsometric permittivity data and fits for the materials used in the multiwavelength metalens design.}
\end{figure}

Plots of the experimental material data that was used during simulation are in Figure \ref{fig:material data}. Dielectric permittivity values for aSi:H were obtained via ellipsometry from hydrogenated amorphous silicon thin films deposited by plasma enhanced chemical vapor deposition (PECVD). PMMA dielectric permittivity was retrieved from spin-coated thin film (E-beam resist PMMA 950K from Allresist) and SiO$_2$ from 1 mm thick fused silica substrate measured by Sentech SE850 ellipsometer.  For the RCWA simulations, the complex permittivity was extrapolated from this data by fitting to a basic Drude-Lorentz dispersion model given by:
\begin{equation}
	\varepsilon(\omega) = \varepsilon_{\infty} + \sum_{p=1}^P\varepsilon_p(\omega),
\end{equation}
where
\begin{equation}
	\varepsilon_p(\omega) = \frac{a_{p,0}+ia_{p,1}(-i\omega)}{b_{p,0}+b_{p,1}(-i\omega)-b_{p,2}\omega^2}.
\end{equation}
Here, $\varepsilon_{\infty}$ and the $a$ and $b$ values are the free fitting parameters and the number of oscillators, $P$ was set to three. Plots of these fits used in extrapolation are also shown in Figure \ref{fig:material data}. For the FDTD simulations, experimental data were loaded into the Lumerical software and its internal fitting functions used to extrapolate over the required frequency range.

\newpage
\bibliography{supp.bib}% Produces the bibliography via BibTeX.